\documentclass[11pt,preprint]{aastex}
\usepackage{apjfonts}
\usepackage{epsfig}
\usepackage{lscape,graphicx}
\usepackage{url}
\usepackage{natbib}
\usepackage[nolists,nomarkers]{endfloat}

\defcitealias{gcm08}{GCM08}
\defcitealias{g06a}{G06}
\defcitealias{cmg+06}{CD06}

\shorttitle{Kinematics of Anticenter Stream Tidal Debris in SA 76}
\shortauthors{Carlin et al.}

\begin{document}

\title{Kinematics in Kapteyn's Selected Area 76: Orbital Motions
Within the Highly Substructured Anticenter Stream}

\author{
Jeffrey L. Carlin\altaffilmark{1,2}, 
Dana I. Casetti-Dinescu\altaffilmark{3,4}, 
Carl J. Grillmair\altaffilmark{5},
Steven R. Majewski\altaffilmark{1}, \& 
Terrence M.Girard\altaffilmark{3}}

\altaffiltext{1}{Department of Astronomy, University of Virginia,  P.O Box 400325, Charlottesville, VA 22904-4325 (jcarlin, srm4n@virginia.edu)}
\altaffiltext{2}{Visiting Astronomer, Kitt Peak National Observatory, National Optical Astronomy Observatory, which is operated by the Association of Universities for Research in Astronomy (AURA) under cooperative agreement with the National Science Foundation.}
\altaffiltext{3}{Astronomy Department, Yale University, P.O. Box 208101,
New Haven, CT 06520-8101 (dana@astro.yale.edu, girard@astro.yale.edu)}
\altaffiltext{4}{Astronomical Institute of the Romanian Academy, Str.
Cutitul de Argint 5, RO-75212, Bucharest 28, Romania}
\altaffiltext{5}{Spitzer Science Center, 1200 E. California Blvd., Pasadena, CA 91125 (carl@ipac.caltech.edu)}

\begin{abstract}

We have measured the mean three-dimensional kinematics of stars in
Kapteyn's Selected Area (SA) 76 ($l$~=~209.3$\arcdeg$,
$b$~=~26.4$\arcdeg$) that were selected to be Anticenter Stream (ACS)
members on the basis of their radial velocities, proper motions, and
location in the color-magnitude diagram.  From a total of 31 stars
ascertained to be ACS members primarily from its main sequence
turnoff, a mean ACS radial velocity (derived from spectra obtained
with the Hydra multi-object spectrograph on the WIYN 3.5m
telescope\footnote{The WIYN Observatory is a joint facility of the
University of Wisconsin-Madison, Indiana University, Yale University,
and the National Optical Astronomy Observatory.}) of $V_{\rm helio} =
97.0 \pm 2.8$ km s$^{-1}$ was determined, with an intrinsic velocity
dispersion $\sigma_{o} = 12.8 \pm 2.1$ km s$^{-1}$.  The mean absolute
proper motions of these 31 ACS members are $\mu_{\alpha}$ cos $\delta
= -1.20\pm0.34$ mas yr$^{-1}$, and $\mu_{\delta} = -0.78\pm0.36$ mas
yr$^{-1}$.  At a distance to the ACS of 10$\pm$3 kpc, these measured
kinematical quantities produce an orbit that deviates by
$\sim$30$\arcdeg$ from the well-defined swath of stellar overdensity
constituting the Anticenter Stream in the western portion of the Sloan
Digital Sky Survey footprint.  We explore possible explanations for
this, and suggest that our data in SA 76 are measuring the motion of a
kinematically cold sub-stream among the ACS debris that was likely a
fragment of the same infalling structure that created the larger ACS
system.  The ACS is clearly separated spatially from the majority of
claimed Monoceros ring detections in this region of the sky; however,
with the data in hand, we are unable to either confirm or rule out an
association between the ACS and the poorly-understood Monoceros
structure.

\end{abstract}


\section{Introduction}
\subsection{The Anticenter Stream}

One of the many stellar overdensities detected in Sloan Digital Sky
Survey (SDSS) data is the so-called ``Anticenter Stream'' (ACS)
unveiled by \citet[hereafter G06]{g06a} using matched-filter star
counts.  The ACS feature is seen as a well-defined swath of stellar
excess near the Galactic anticenter at moderately low latitudes
18$\arcdeg \lesssim b \lesssim$ 35$\arcdeg$ (at roughly constant right
ascension, $\alpha_{2000} \approx$ 125$\arcdeg$), spanning the full
$\sim$65$\arcdeg$ declination coverage of the SDSS database in this
region.  \citetalias{g06a} discovered distinct, well-separated narrow
``tributaries'' or sub-streams within the broad ACS stream; these
tributaries are thought to be dynamically distinct components among
the remnants of a tidally disrupted dwarf galaxy.  \citet[hereafter
GCM08]{gcm08} measured radial velocities in two fields along the ACS,
and identified the radial velocity signature of the stream.  The
relatively large measured velocity dispersion for the northernmost of
the two fields (ACS-B; $\sigma_V \sim 15$ km s$^{-1}$) is consistent
with the dispersion for a tidal remnant of a disrupted dwarf galaxy,
but may also result from the sampling of more than one of the apparent
cold sub-streams within the broader ACS.  For the second field, ACS-C,
about 23$\arcdeg$ South of ACS-B, \citetalias{gcm08} found a
dispersion of only $\sim$5 km s$^{-1}$, suggesting that their data in
this field sample only one of the kinematically cold ``tributaries''
making up the larger system.  The mean velocities in each field,
combined with SDSS/USNO-B proper motions and the locations of stream
overdensities, were used by \citetalias{gcm08} to fit an orbit and
show that the ACS corresponds to debris lost from an object on a
low-inclination, nearly circular, prograde orbit.  This
\citetalias{gcm08} orbit succeeds in reproducing not only the broad
swath cut by the ACS stream in the SDSS footprint, but also suggests
that the ``Eastern Banded Structure'' (EBS) pointed out by
\citetalias{g06a} at ($\alpha$, $\delta$)$_{2000} \sim
(134\arcdeg,3.4\arcdeg)$ is associated with the ACS as debris from a
subsequent (or prior) orbital wrap of the same system.

\begin{figure}%[!htp]
%\epsscale{1.2} %increases the size to fit column in 2-column format
\plotone{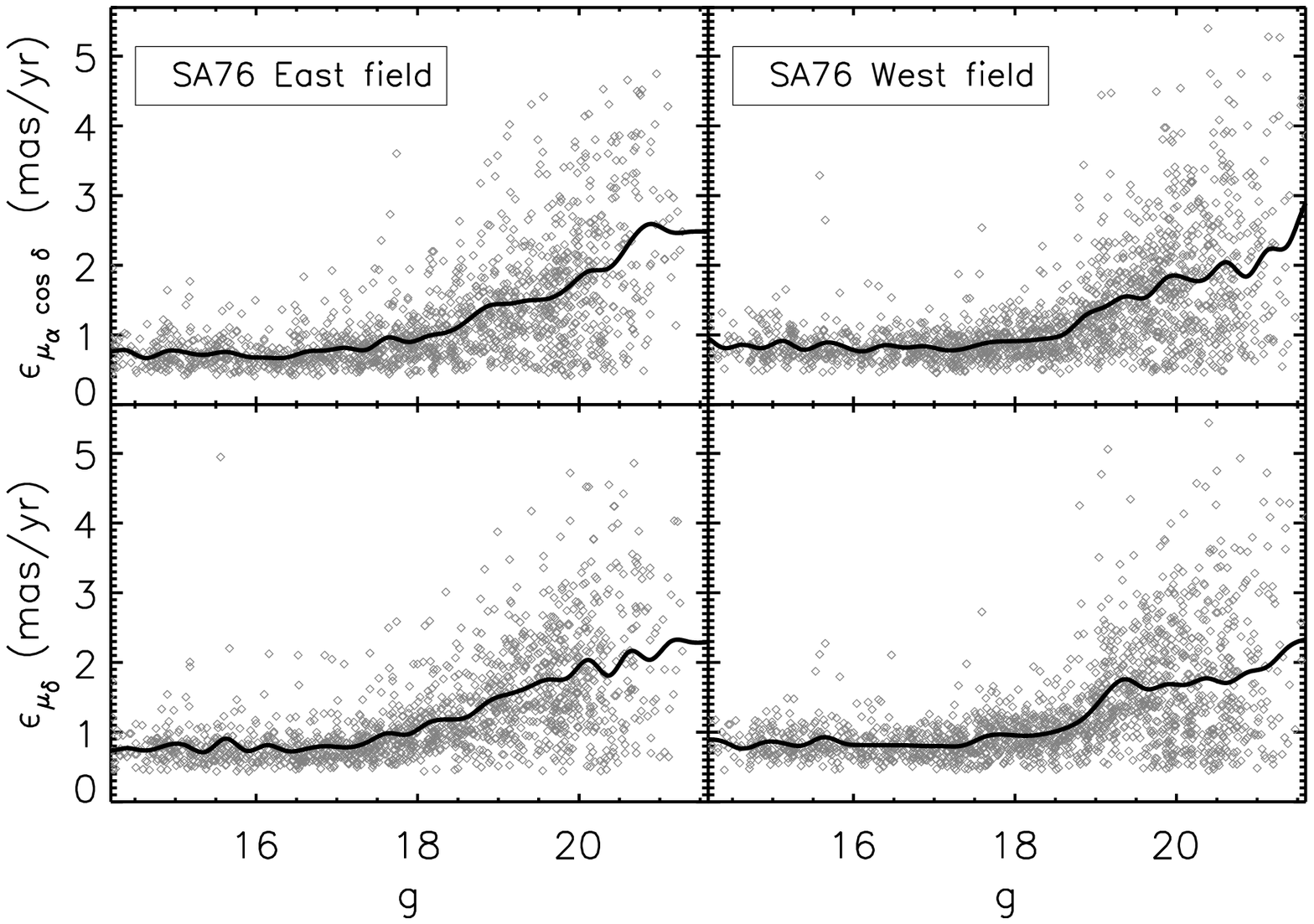}
\caption{Proper motion errors as a function of $g$ magnitude for the
east and west fields.  The solid lines show a moving median in
bins of 0.25-magnitude width. Most well-measured stars have proper
motion uncertainties of 1-2 mas yr$^{-1}$ in each
direction.\label{fig:pm_errs}}
\end{figure}

\subsection{Previously Studied Stellar Overdensities in the Same Region: Monoceros/GASS}

The stellar overdensity in this region of the SDSS database was
identified by \citet{bze+06} as being associated with the Monoceros
(Mon) Ring, a set of features first reported by \citet{nyr+02},
\citet{yng+03}, \citet{rms+03}, and \citet{iil+03} that seem to form a
low-latitude ringlike structure over a large area near the Galactic
anticenter.\footnote{This low-latitude structure and other seemingly
related features have been christened variously the ``Monoceros
Stream'' \citep{yng+03}, the ``One Ring'' \citep{iil+03}, and GASS
(Galactic Anticenter Stellar Structure; \citealt{rms+03}).  It is not
clear that the distinct, narrow stream seen in background-subtracted,
filtered starcount maps of the SDSS database \citep{g06a} is
physically associated with the much larger purported association of
overdensities comprising Monoceros/GASS.  For convenience, we follow
the naming convention from \citetalias{g06a}, and refer to the narrow
stream structure uncovered by these authors in the Western portion of
the SDSS footprint as the ``Anticenter Stream'', or ACS.  To avoid
conflating possibly distinct structures, all other overdensities that
have been associated with the Monoceros feature in past works will be
referred to in this paper as part of Monoceros (Mon).  We acknowledge
that some or all of these structures may be associated, and if that is
found to be the case, the names might thus be interchangeable.}
However, once the smoothly-varying background has been removed
\citepalias{g06a} the Anticenter Stream appears to be a separate
narrow stream structure, rather than a part of the much larger
Monoceros feature.  Nevertheless, because many detected overdensities
in this region of sky have been attributed to Monoceros, we will
compare our ACS findings to existing Monoceros data.

Based on its narrow radial extent and low velocity dispersion,
\citet{cmr+03} argued that Mon was the remnant of a tidally disrupted
dwarf galaxy.  \citet{iil+03} followed up with photometric detections
spanning $\sim$100$\arcdeg$, and suggested that the apparent ring was
a result of flaring or warping of the Galactic disk.  Evidence of a
rather thin radial extent based on main-sequence width
\citep{yng+03,iil+03} as well as apparent separation from the edge of
the Galactic disk \citep{rms+03} favor an accretion origin for the
ring.  These facts were taken by \citet{cmr+03} along with their
measured radial velocity trend with longitude to be consistent with a
dynamically young tidal stream on a clearly non-circular orbit (though
with low eccentricity).  Low-latitude, torus-like features are
produced by accretion of satellites on nearly-circular orbits roughly
coplanar with the Galactic disk in $\Lambda$ CDM simulations
\citep{ans+03,bj05}, and are also visible in the extragalactic systems
imaged by \citet{mgc+10}.  \citet{cmr+03} also noted that a few
Galactic globular clusters may be associated with the Monoceros ring
in both position and radial velocity -- a point that was further
developed by \citet{fmc+04} to include 11 old open clusters that are
also near the Monoceros plane.  \citet{cli+05,cll+07,cll+08} have
continued to map the extent and stellar populations of the ring, and
have detected overdensities both above and below the Galactic plane,
ruling out a warp or flare of the disk as the origin of the feature.
However, recent simulations \citep{kbz+08,ybc+08} have shown that
interactions of massive subhalos with the Galactic disk can form
ringlike features that result from excitation of thin/thick disk
stars.  On the other hand, recent high-resolution abundance analysis
by \citet{cmc+10} has shown that Monoceros stars have chemical
abundance patterns similar to those of Milky Way dwarf spheroidals
(dSphs) and unlike stars from the outer Milky Way disk (Chou et al.,
{\it in prep.}), bolstering the case for a tidal origin for Mon.
Further detailed chemodynamical studies of stream stars are clearly
needed to assess the origin of this feature and its possible
association with the ACS.

Monoceros has been modeled as debris from a tidally disrupting dwarf
galaxy by both \citet{pmr+05} and \citet{mib+04}, both of whose models
were constrained so that the progenitor reproduces the putative Canis
Major dwarf galaxy (which was discovered by \citealt{mib+04}; but
cf. \citealt{rms+06,mzg+06,mvc+06,lmz+07} for alternative explanations
of this apparent stellar overdensity).  The \citet{pmr+05} models
incorporated all detections of suspected Monoceros debris known at the
time, and reproduced the known structure with a disrupting dwarf
galaxy on a low-inclination, nearly-circular prograde orbit.
Subsequent mapping of the extent and stellar populations of the Ring
have yet to be incorporated into the models, and to date only a few
kinematical constraints have been derived (e.g.,
\citealt{cmr+03,yng+03}).  One of the only proper motion results for
Monoceros debris, measured by \citet{ccg+08} in SA 71, produced an
orbit in agreement with the \citet{pmr+05} determination.  This SA 71
study illustrated the difficulty of studying low-latitude features --
large numbers of accurate proper motions supplemented by radial
velocities were necessary to distinguish the kinematical signature of
Mon from the overlapping kinematics of Galactic populations.

\begin{figure}%[!htp]
%\epsscale{1.2} %increases the size to fit column in 2-column format
\plotone{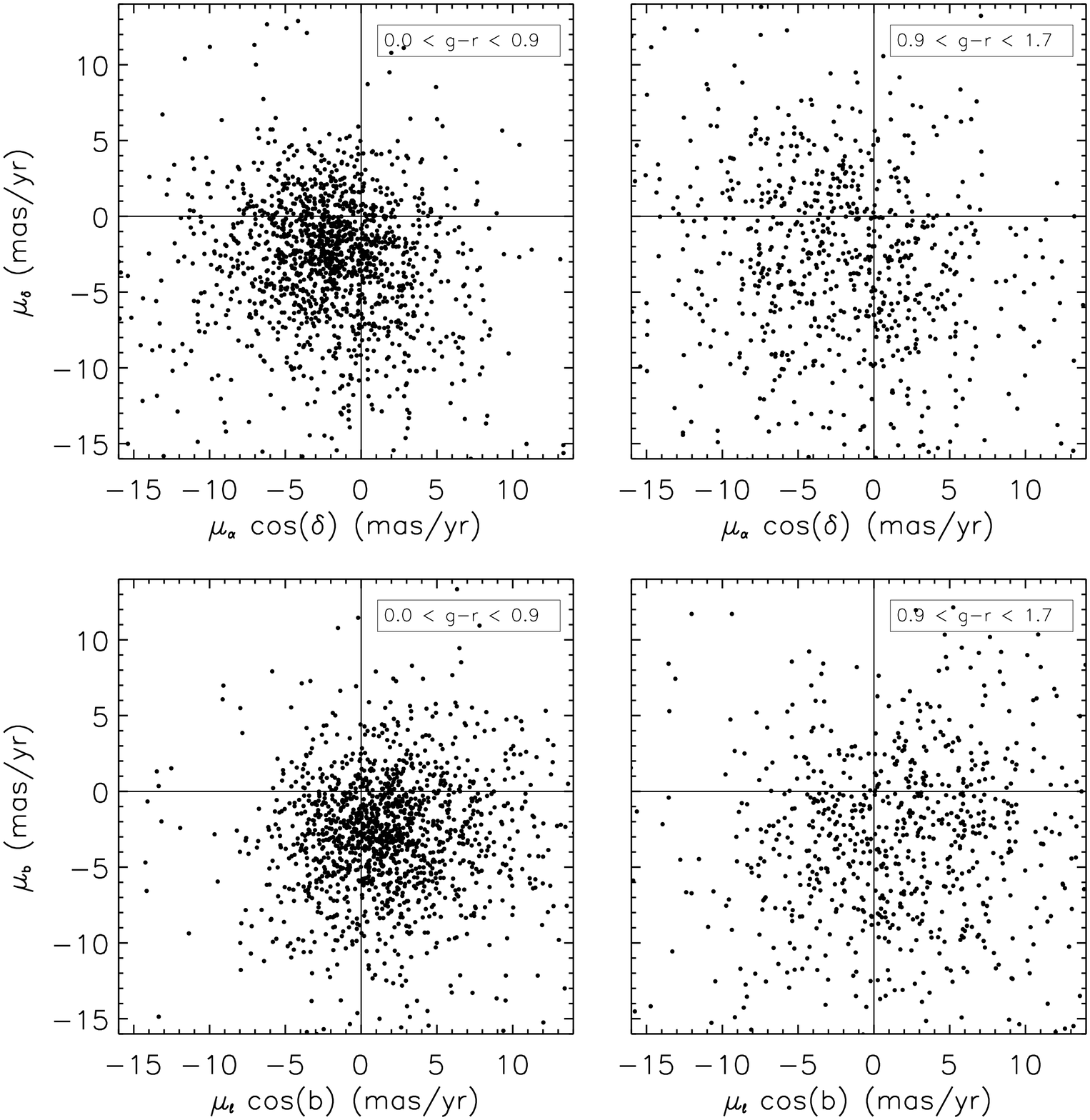}
\caption{Proper motions of all stars in SA 76 measured on at least
four plates.  The upper two panels show proper motions in equatorial
coordinates, and the bottom panels along Galactic coordinates.  These
are divided into a red ($0.9 < g-r < 1.7$) sample made up of primarily
foreground Milky Way dwarfs, and blue ($0.0 < g-r < 0.9$) stars, which
are mostly Milky Way thick disk and halo stars, along with ACS debris.
The more distant thick disk, halo, and ACS stars show more tightly
clumped proper motions. \label{fig:vpd_all}}
\end{figure}

\begin{figure}%[!htp]
%\epsscale{1.2} %increases the size to fit column in 2-column format
\epsscale{0.9}
\plotone{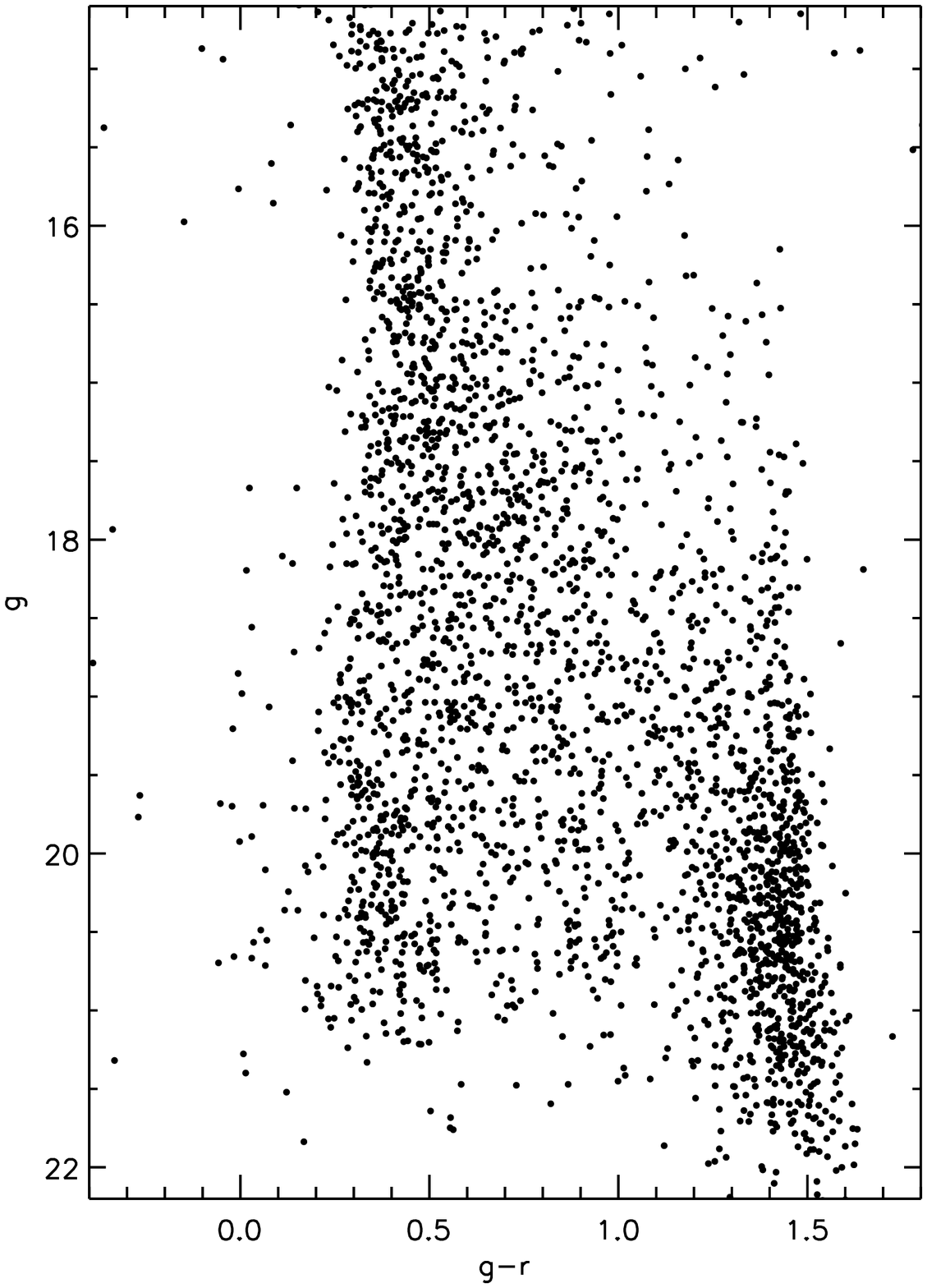}
\caption{SDSS color magnitude diagram of all stars with well-measured
proper motions.  The proper motion catalog is $\gtrsim68\%$ complete
at $g = 20.0$.  Anticenter Stream debris is noticeable as an
overdensity at faint ($g > 19$), blue ($g-r<0.5$)
colors.\label{fig:cmd_all}}
\end{figure}

\subsection{Goals of This Paper}

Here, we present a follow-up to the \citetalias{gcm08} study,
supplementing the radial velocities from that previous work with
substantially more measurements from WIYN+Hydra spectroscopy as well
as accurate proper motions in one of the fields studied there.  This
field, ACS-C, was originally selected because it coincides with SA 76,
one region from the deep proper motion study in Kapteyn's Selected
Areas (SAs) by \citet{m92} and \citet[hereafter CD06]{cmg+06}.  From
the newly-obtained spectra, additional likely radial velocity members
of the ACS are identified and used to derive the mean absolute proper
motion (and thus 3-D kinematics) of stream stars along this line of
sight.  This represents one of the most precise absolute proper
motions yet derived for a Milky Way stellar tidal stream.

The organization of this paper is as follows: \S 2 gives an overview
of the proper motion and spectroscopic data for SA 76 and their
handling, including details of both the astrometric and spectroscopic
data reduction.  The selection of Anticenter Stream members is
detailed in \S 3.  Initially, candidates were selected based on broad
radial velocity criteria.  Additional culling of the sample was
performed based on reduced proper motions.  In \S 4, we present
metallicities for stars selected to be ACS members, comparing our
results to the expected [Fe/H] distribution of foreground stellar
populations selected from the Besan\c{c}on galaxy model.  Section 5
discusses the measured kinematics based on our final sample of
Anticenter Stream debris candidates.  The mean radial velocity and
absolute proper motions yield a low-inclination, nearly-circular orbit
for ACS debris in SA 76.  The mean 3-D space motion we find for SA 76
ACS candidates is oriented at a $\sim30\arcdeg$ angle to the
prominently visible Anticenter Stream.  Some possible explanations for
this are discussed in \S 6 -- we believe that the motions we have
measured represent a kinematically distinct substream from within the
more extensive ACS system.  Finally, we compare our results to known
characteristics of the Monoceros ring in \S 7.  From the available
data, the ACS appears to be unrelated to Monoceros, but this remains
unclear.

% refer using: Table~\ref{tab:obs_tab}
\begin{table}%[!htp]
\begin{center}
\caption{Summary of WIYN+Hydra Spectroscopic Observations in SA 76\label{tab:obs_tab}}
\begin{tabular}{cccc}
\\
\tableline
\\
\multicolumn{1}{c}{Date} & \multicolumn{1}{c}{Exposures} & \multicolumn{1}{c}{N$_{\rm stars}$} & \multicolumn{1}{c}{Mag. limit\tablenotemark{a}} \\
 & (seconds) & & \\
\tableline
\\
 Dec 2006\tablenotemark{b} & 3 x 1800 & 66 & 18.5 \\ 
 Feb 2007\tablenotemark{b} & 9 x 2400 & 61 & 20.1 \\
 Dec 2007 & 6 x 2700 & 57 & 20.0 \\
 Nov 2008 & 8 x 1800 & 59 & 20.1 \\
\multicolumn{2}{c}{\bf ... TOTAL .................... } & {\bf 224}\tablenotemark{c} & \\
%\\
\tableline
\end{tabular}
\tablenotetext{a}{SDSS g magnitudes.}
\tablenotetext{b}{These observations were reported in GCM08 as their field ``ACS-C''.}
\tablenotetext{c}{Total number is less than the sum of targets from individual runs because some stars were oberved on multiple runs.}
\end{center}
\end{table}

\section{The Data}
\subsection{Photometry and Proper Motions}

Photometry from SDSS Data Release 5 (DR5; \citealt{aaa+07}) was used
for most analysis in this work (DR7 was used for the starcount
maps). Throughout this contribution, quoted magnitudes are measured
SDSS magnitudes (i.e., {\it not} dereddened or extinction corrected);
the mean reddening along this line of sight from the dust maps of
\citet{sfd98} is $E(B-V) = 0.04$.  The main sequence of the ACS is
readily distinguished at $g\gtrsim18.5$ over the large area in which
\citetalias{g06a} mapped the stream.  \citetalias{gcm08} used
filtered, background-subtracted number counts (see \citealt{g09} for
details of the matched-filter technique) to identify regions of
highest stream density from which to select spectroscopic targets
along the ACS.  Analysis in
\citetalias{gcm08} focused on two widely-spaced fields within the ACS:
ACS-B at $(\alpha,\delta)_{2000} = (124\arcdeg,37.5\arcdeg)$ and ACS-C
at $(\alpha,\delta)_{2000} = (125\arcdeg, 14.7\arcdeg)$.  The first of
these, ACS-B, was selected because of its relatively high
matched-filter stellar density, while the latter field, ACS-C at
$(l,b)= (209.3\arcdeg,26.4\arcdeg)$, was chosen because it overlaps SA
76, one of the regions studied in the deep proper motion survey first
described in \citep{m92} (see also \citetalias{cmg+06}).  We note that
SA 76 is offset slightly east of the highest-density portion of the
stream, but is still in a region of elevated stellar density.

\begin{figure}%[!htp]
\plotone{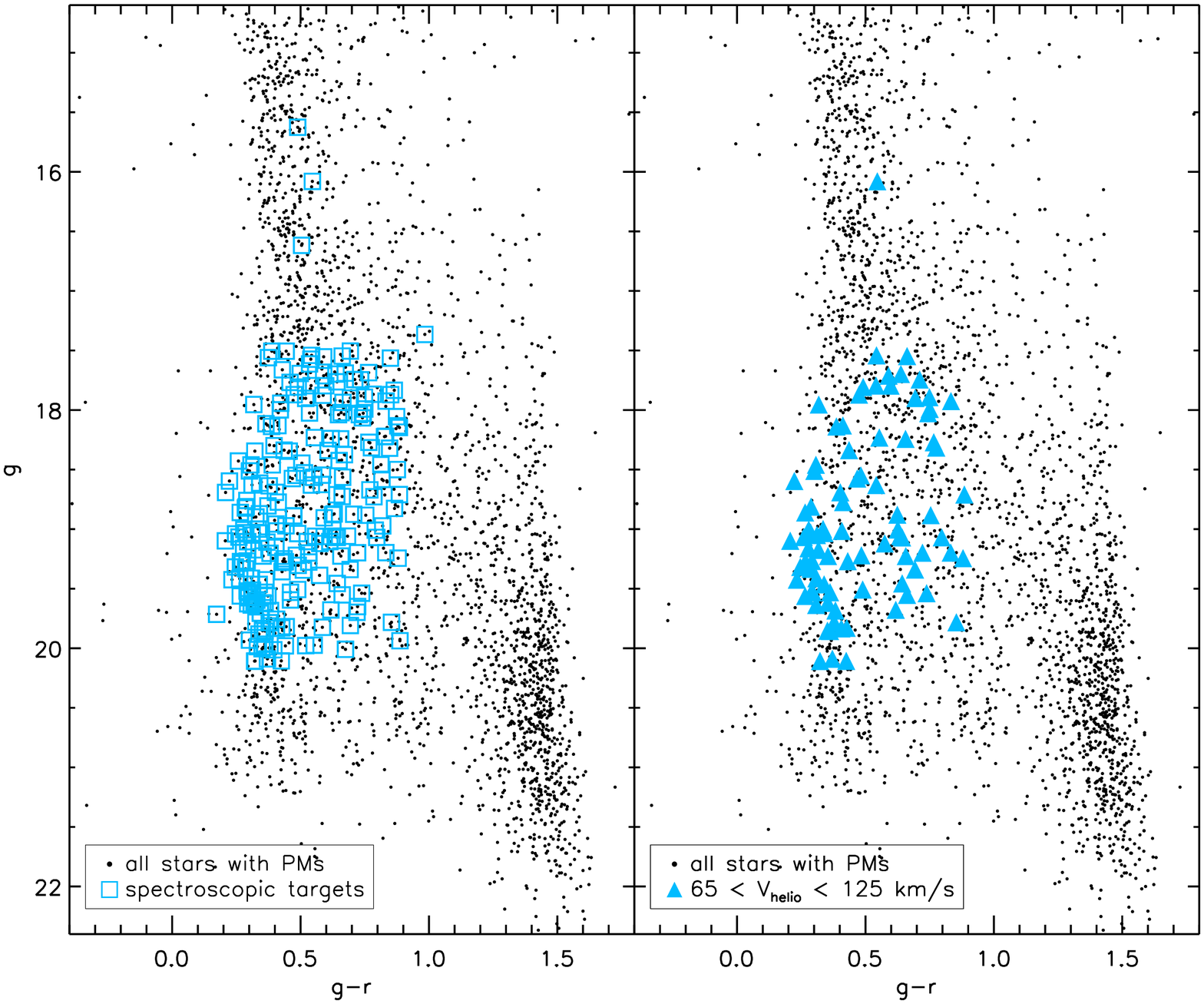}
\caption{Color magnitude diagrams of all stars with well-measured
proper motions, highlighting the spectroscopically observed samples.
{\it Left panel:} Open squares show all stars observed
spectroscopically (see Section 2.1 for details on target
selection). {\it Right panel:} Solid triangles represent stars within
the initial 65 $< V_{\rm helio} <$ 125 km s$^{-1}$ velocity selection
discussed in Section 3.  The majority of these selected candidates are
confined to an apparent upper main sequence at $g > 18.5$, $0.2 <
(g-r) < 0.5$. \label{fig:cmd_targetselect}}
\end{figure}

\begin{figure}%[!htp]
\plotone{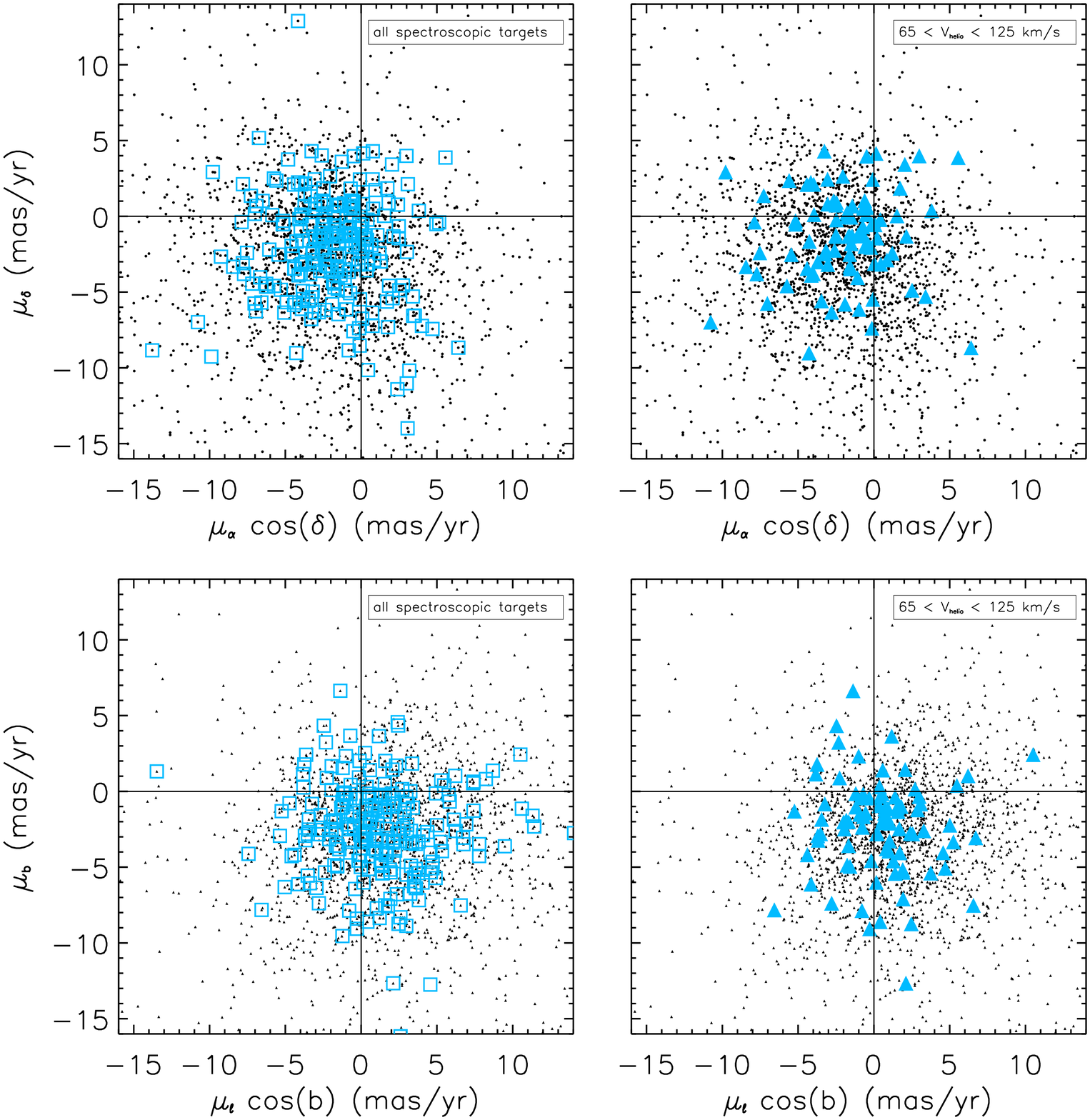}
\caption{All well-measured proper motions in SA 76, highlighting the
same spectroscopically observed samples as
Figure~\ref{fig:cmd_targetselect}. {\it Left panel:} Open squares show
all stars observed spectroscopically. {\it Right panel:} Solid
triangles represent stars within the initial 65 $< V_{\rm helio} <$
125 km s$^{-1}$ velocity selection.  Most candidates thus selected are
clumped more tightly than the overall spectroscopic
sample.\label{fig:vpd_targetselect}}
\end{figure}

SA 76 is one field of $\sim$50 from the Mt.~Wilson 60-inch
telescope-based part of the proper-motion survey described by
\citetalias{cmg+06}; details concerning the data reduction process and
the derivation of proper motions are presented there. Here, we briefly
outline the procedure. The photographic plates were taken at three
different epochs: the modern epoch consists of plates taken between
1996 and 2000 with the Las Campanas du Pont 2.5~m telescope, the
intermediate epoch consists of Palomar Observatory Sky Survey plates
(POSS-I) taken in the early 1950s with the Palomar Schmidt 1.2~m
telescope, and the old epoch consists of two plates taken in 1909 and
1912 at the Cassegrain focus of the Mount Wilson 1.5~m telescope. The
modern and old plates were digitized with the Yale PDS
microdensitometer.  Analysis for the POSS-I plates used scans done by
both the Space Telescope Science Institute (the Digitized Sky Survey)
and the US Naval Observatory, which were then processed at Yale to
obtain more accurate positions than provided by the USNO
catalogs. Typically, each SA field covers $40\arcmin\times40\arcmin$,
an area constrained by the du Pont plates.  Two sets of du Pont plates
offset by $\sim 20\arcmin$ in R.A. (to match the Mt.~Wilson plate
centers) were taken for SA 76 in 1996 and 1998. Each set includes one
blue (IIIa-J + GG385) and one visual (IIIa-F + GG495) plate.  Three
overlapping fields from POSS-I were used, each including one red
(103a-D + RP2444) and one blue (103a-O, no filter) plate.  Proper
motions were measured separately for the east and west fields of SA
76, as given by the two du Pont sets of plates. Only the du Pont
plates were divided into east and west fields -- although these east
and west plate centers were chosen to match those of the older plates,
the coarser plate scale of the Mt. Wilson plates (26 arcsec mm$^{-1}$
vs. 10.9 arcsec mm$^{-1}$ for the du Pont plates) means that each
single Mt. Wilson plate covers the entire area of the combined du Pont
fields. The correction to absolute proper motions was defined by 97
galaxies in the east field and by 144 galaxies in the west field and
applied separately to each field.  The uncertainty in the correction
to absolute proper motions is between 0.41 and 0.45 mas~yr$^{-1}$ in
each dimension for both the east and west fields. Finally, the two
data sets were combined by finding a weighted average absolute
proper-motion for well-measured stars in the overlapping region of the
two fields. Proper-motion uncertainties per star range between 1-2
mas~yr$^{-1}$ for well-measured stars to $\sim 4$ mas~yr$^{-1}$ at the
faint limit of the survey.  The uncertainties are shown as a function
of magnitude for all stars in the east and west fields in
Figure~\ref{fig:pm_errs} -- the solid lines represent a moving median
value (in 0.25-magnitude bins) for proper motion error as a function
of magnitude for each field.

Figure~\ref{fig:vpd_all} shows absolute proper motions (PMs) in the
final combined SA 76 data set for all stars measured on at least four
plates spanning at least two of the three widely-separated mean
epochs.  The proper motions are divided into a red ($0.9 < g-r < 1.7$)
and blue ($0.0 < g-r < 0.9$) sample to highlight kinematical
differences between the predominantly nearby M dwarfs in the red
sample and the more tightly-clumped (in PMs) blue sample comprised
mostly of thick disk and halo stars (note that the distribution is not
centered on, or symmetric about, zero proper motion -- this is mostly
due to the Solar motion, and likely also due a contribution from the
ACS).  By comparison to the SDSS photometric catalog, the
proper-motion catalogs are found to be $82\%$ and $68\%$ complete at
$g = 20.0$ for the west and east fields, respectively.

An SDSS color-magnitude diagram (CMD) for all stars with well-measured
proper motions (i.e. measured on at least 5 plates from at least 2
epochs) in SA 76 is seen in Figure~\ref{fig:cmd_all}.  The main
sequence turnoff (MSTO) of a distant, metal-poor population is
apparent as an overdensity at $g\gtrsim$19, $(g-r)\sim$0.2-0.5, with
the main sequence extending down beyond the limits of the survey.  The
proper motions extend $\sim$1.5 magnitudes fainter than this MSTO
feature.

\subsection{Radial Velocities}
\subsubsection{Sample Selection}

As seen in \citetalias{g06a}, the MSTO feature of the ACS is prominent
in SDSS CMDs along the stream.  We thus focus our selection of stars
for followup spectroscopy within the faint (18.5$ < g < $20.1)
magnitude and blue (0.2$ < g-r < $0.6) color ranges of the turnoff.
As in \citetalias{gcm08}, we used preliminary proper motions from the
\citetalias{cmg+06} survey to constrain spectroscopic target selection
to only those stars within the MSTO region that also have relatively
small PMs ($|{\mu}| < 10$ mas yr$^{-1}$ in each dimension) expected
for the distant Anticenter Stream.  Additional targets outside the
narrow color, magnitude, and proper motion selection regions were
included at lower priority, in order to fill all of the spectrograph
fibers.  Furthermore, one fiber configuration of stars at bright
magnitudes and colors consistent with the red giant branch (RGB) of
the ACS population was observed as a backup target during marginal
weather conditions.  All targets with spectra having sufficient
signal-to-noise that velocities could be derived are shown in the CMD
in the left panel of Figure~\ref{fig:cmd_targetselect}, as well as the
proper-motion vector point diagram (VPD) of
Figure~\ref{fig:vpd_targetselect} (left panel).  Care was taken not to
be too restrictive in the color and proper-motion selections, since we
had no a priori knowledge of the exact location of ACS debris in
either of these dimensions.

\subsubsection{Observations}

Radial velocities (RVs) were derived from spectroscopic data obtained
over the course of four observing runs with the WIYN 3.5-m telescope
between Dec. 2006 and Nov. 2008 (Table~\ref{tab:obs_tab}). The Hydra
multi-fiber spectrograph was used in two similar configurations. The
Feb. 2007 observing run used the 600@10.1 grating in first order with
the red fiber cables at a wavelength center of 5400\AA~, yielding
wavelength coverage over 4000--6800\AA~at a dispersion of 1.397
\AA$~$pix$^{-1}$, and a spectral resolution of 3.35 \AA~.  These data
are discussed in GCM08 as the ``ACS-C'' spectra.  During the
Dec. 2006, Dec. 2007, and Nov. 2008 runs, a similar configuration was
utilized (same grating, etc.), but centered slightly redward,
providing wavelength coverage $\lambda$ = 4400--7200 \AA~at the same
dispersion and resolution.  The general spectral region was selected
to include the H$\beta$, Mg triplet, Na D, and H$\alpha$ spectral
features.  We further note that the Nov. 2008 observing run occurred
after the WIYN Bench Spectrograph Upgrade, which included the
implementation of a new collimator into the Bench configuration, as
well as a new CCD that delivers greatly increased throughput.  Each
Hydra configuration was exposed multiple times to enable cosmic ray
removal.  Exposure times, number of stars targeted, and the limiting
magnitude of stars in each observed configuration can be found in
Table~\ref{tab:obs_tab}.  The December 2006 observations were limited
to bright targets due to weather (we were only able to observe briefly
between periods of snow and high humidity), while the February 2007
observations were beset by high winds and poor ($\sim2.5\arcsec$)
seeing.  Good seeing ($\sim0.6\arcsec$) and clear skies on both the
December 2007 and November 2008 observing runs allowed observation of
the majority of faint targets within the selection.  Typically 60-70
targets were placed on Hydra fibers, with the remaining 15-20 fibers
placed on blank sky regions to allow for accurate sky subtraction.

\begin{figure}%[!htp]
\epsscale{0.8}
\plotone{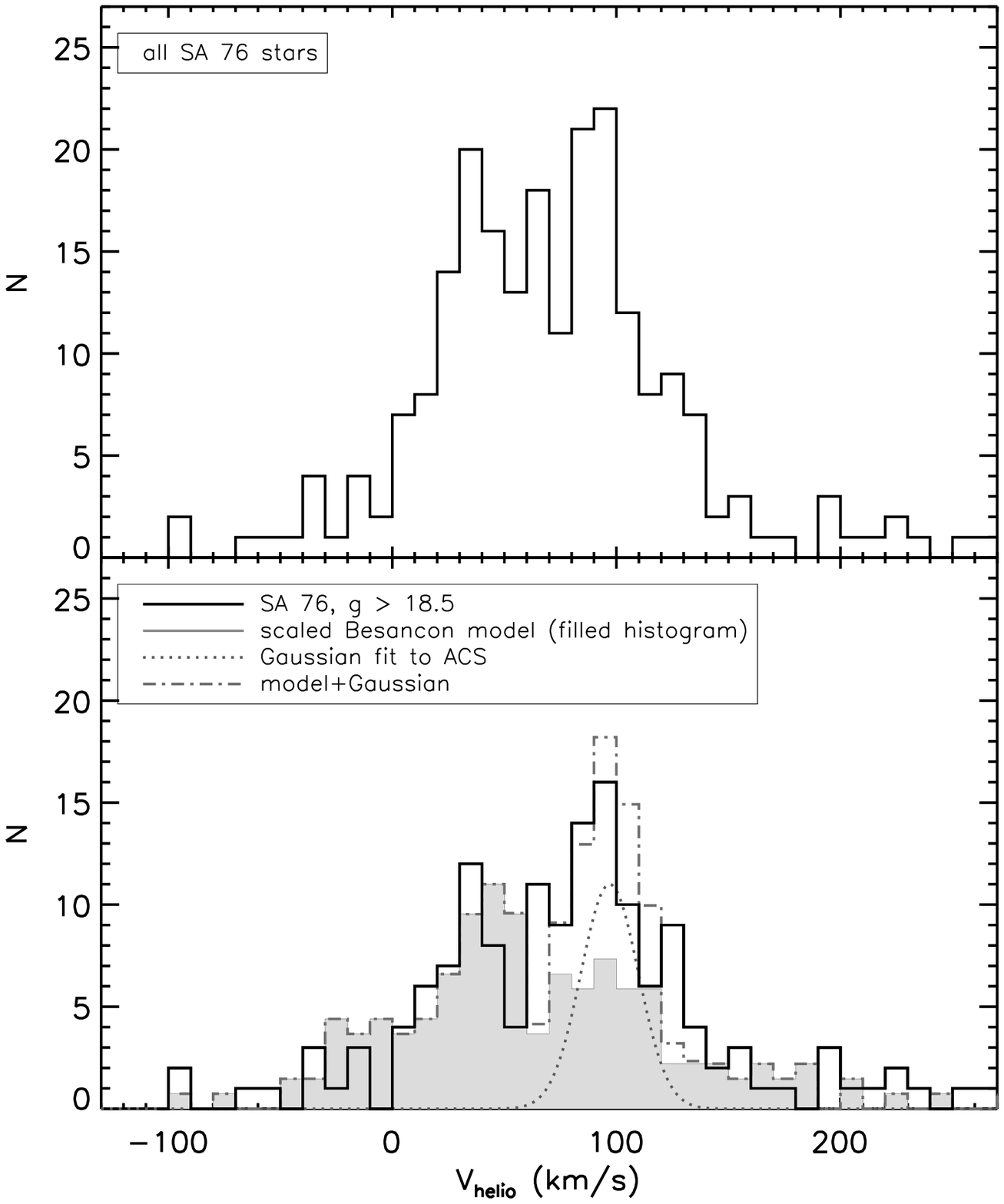}
\caption{{\it Upper panel:} Measured heliocentric radial velocities of
all observed targets in SA 76.  Two prominent peaks are visible at
$V_{\rm helio} \sim 35$ and $V_{\rm helio} \sim 90 $ km s$^{-1}$.  {\it
Lower panel:} Measured heliocentric velocities for all targets in SA
76 having magnitudes $g >$ 18.5. The filled histogram shows the
expected velocities from a scaled sum of five Besan\c{c}on model
queries along the SA 76 line of sight, selected within the same
magnitude and color ranges as the ACS candidates.  A Gaussian with the
maximum likelihood results for the mean velocity ($V_{\rm helio} =
97.0$ km s$^{-1}$) and dispersion ($\sigma_{0} = 12.8$ km s$^{-1}$) of
the Anticenter Stream is overplotted as a dotted curve.  The sum of
the Besan\c{c}on model (representing Galactic stellar populations, and
scaled so that the final dot-dashed histogram matches the number of
stars observed spectroscopically) and the best-fit Gaussian (i.e. ACS
members) is shown as a dot-dashed histogram.
\label{fig:vhel_hist}}
\end{figure}

\begin{figure}%[!htp]
\epsscale{0.9}
\plotone{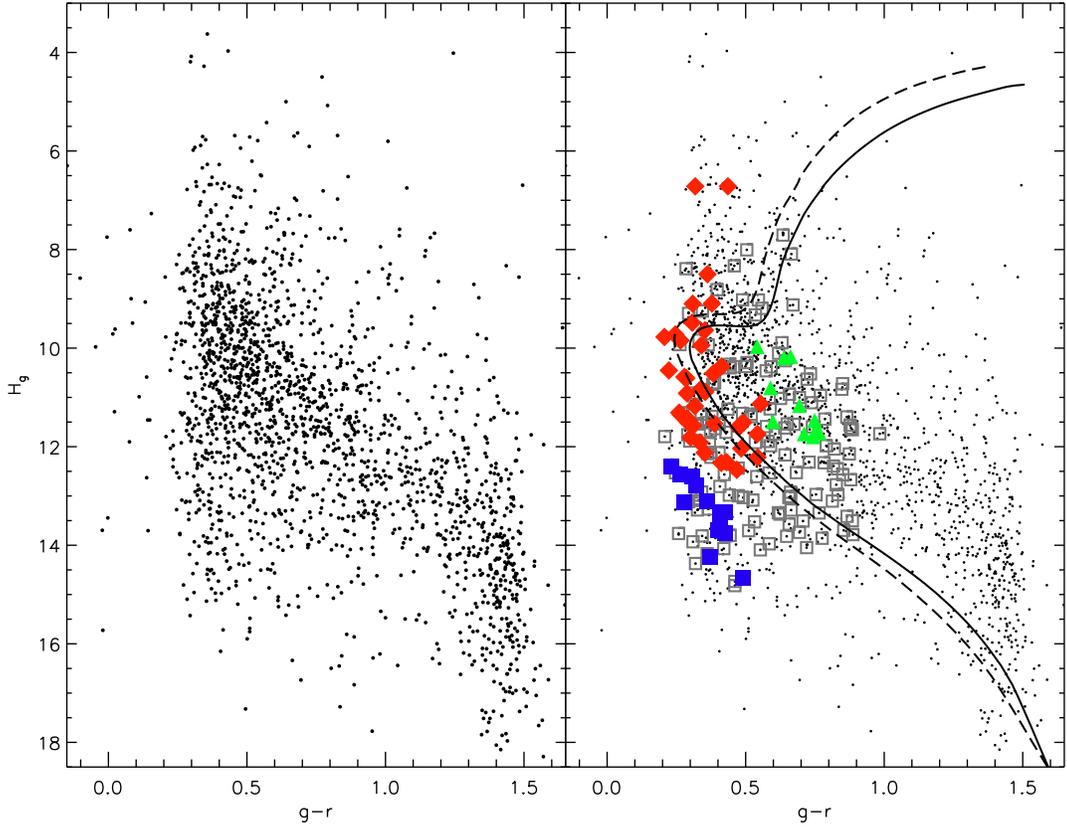}
\caption{Reduced proper motion diagram (RPMD) for all well-measured
stars in SA 76, where $H_g \equiv g + 5$ log $\mu + 5$.  The left
panel shows all well-measured objects.  The swath of stars running
across the diagram from $g-r \sim 0.4, H_g \sim 9$ to $g-r \sim 1.4,
H_g \sim 15$ consists mainly of nearby Milky Way disk stars, with the
vertical feature at $g-r \sim 1.4$ made up of local M-dwarfs.  The
extension to higher values of $H_g$ at $g-r \sim 0.4$ (i.e., below the
disk stars) represents either a population at larger tangential
velocity than the disk stars or metal-poor Galactic subdwarfs.  In the
right panel, all spectroscopic targets are depicted with open (grey)
squares, with solid symbols representing stars within the initial RV
selection 65 $< V_{\rm helio} <$ 125 km s$^{-1}$.  For reference, two
10 Gyr ridgelines from \citet{dcj+08} at a distance of 10 kpc, with
tangential velocity of 70 km s$^{-1}$, and [Fe/H] = -1.3 and -0.9
(solid and dashed lines, respectively) are shown. These ridgelines
have also been corrected for the mean reddening ($E(B-V) = 0.04$
according to the \citealt{sfd98} maps) along the line of sight. Stars
in the RV-selected sample at red (0.5 $< g-r <$ 0.9) colors having
$H_g$ inconsistent with the expected ACS main sequence and giant
branch were removed from the sample; these are shown as filled green
triangles.  The remaining RV-selected candidates were separated into
two groupings that resemble MSTO features at blue (0.2$ < g-r < $0.5)
colors and reduced proper motions of 11 $\lesssim H_g \lesssim$ 12.5
and 12.5 $\lesssim H_g \lesssim$ 14.5 -- a ``lower'' (solid blue
squares) and an ``upper'' (filled red diamonds) sample.  These two
samples were examined to determine whether there are two distinct
turnoffs at different tangential velocities in SA 76 (and within the
RV selection).  The lower sample (blue squares) apparently consists of
foreground (or background) contamination by MW halo subdwarfs, and was
removed from the final ACS sample. \label{fig:gr_rpmd}}
\end{figure}

\begin{figure}%[!htp]
%\epsscale{1.2} %increases the size to fit column in 2-column format
\plotone{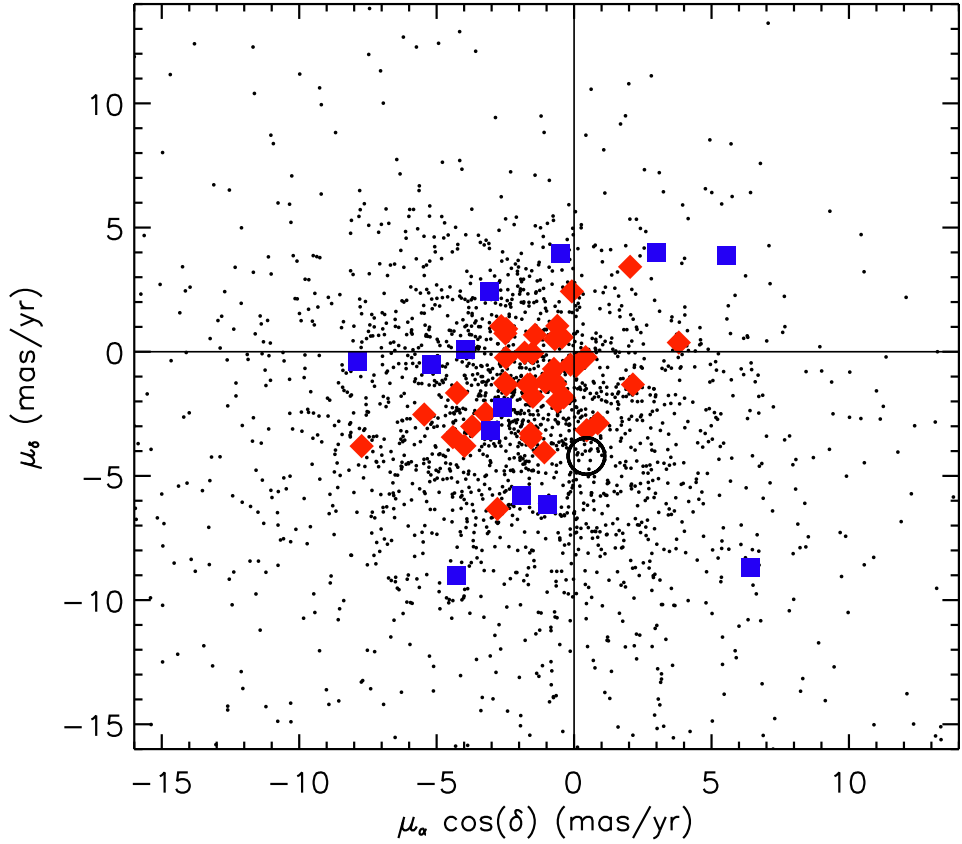}
\caption{Vector point diagram of all well-measured stars in SA 76.
The upper and lower RPMD samples (as defined in
Figure~\ref{fig:gr_rpmd}) are shown as red diamonds and blue squares,
respectively.  The red diamonds (upper RPMD sample) clump tightly in
proper motions, as expected for a distant, common-motion stellar
population.  The lower RPMD sample (blue squares) show large scatter,
suggesting they are Milky Way stars (likely metal-poor halo subdwarfs)
in the foreground (or background) of the ACS rather than a
coherently-moving stellar population.  To select a relatively ``pure''
ACS sample, we removed all ``lower RPMD'' stars. The open circle
represents the solar reflex motion for an object at 10 kpc (i.e., the
heliocentric proper motion one would measure for an object stationary
with respect to the Galaxy) along the SA 76 line of
sight. \label{fig:vpd_rpmdsel}}
\end{figure}

Standard pre-processing was performed on the initial two-dimensional
spectra using the CCDRED package in IRAF,\footnote{IRAF is distributed
by the National Optical Astronomy Observatory, which is operated by
the Association of Universities for Research in Astronomy (AURA) under
cooperative agreement with the National Science Foundation.} and
frames were summed before spectrum extraction.  Extraction of 1-D
spectra and further spectroscopic reduction used the DOHYDRA utilities
(also in IRAF).  CuAr arc lamp exposures were taken at each
configuration; from these, 30--35 emission lines were used to fit the
dispersion solution for each Hydra configuration.  A few radial
velocity standards were targeted on each observing run covering
spectral types from F through early K (both dwarfs and giants), each
through multiple fibers, to yield multiple individual
cross-correlation template spectra.  These RV standard spectra were
first cross-correlated against each other using the IRAF tool FXCOR to
determine the accuracy of the velocities and remove any outliers
(i.e. those that yield unreasonable cross-correlation results due to
some defect, such as a poorly-removed cosmic ray).  Measured
velocities of the RV standards typically agreed with published IAU
standard values to within 1-2 km s$^{-1}$.  Radial velocities for
program stars were derived by cross-correlating all object spectra
against all of the standards taken on the same observing run.  To
maximize the cross-correlation $S/N$ in faint, metal-poor stars, only
the regions around the H$\beta$, Mg triplet, and H$\alpha$ absorption
lines were used for cross-correlation.

Radial velocity uncertainties were derived using the \citet{vmo+95}
method, as described in \citet{mcf+06} and \citet{fmp+06}.  The
Tonry-Davis ratio (TDR; \citealt{td79}) scales with $S/N$, such that
individual RV errors can be calculated directly from the TDR, provided
you have multiple observations of some particular standard star to map
the dependence.  Where possible, we have used this technique, but for
the Dec. 2006 observing run, only a total of four RV standard spectra
were taken.  Thus, the RV uncertainty for the Dec. 2006 object stars
is the standard deviation of the results from cross-correlation
against these four standards.  Typical RV uncertainties for individual
measurements were $\sigma_{V} \approx 5-10$ km s$^{-1}$, with most
spectra having $S/N\sim$15--20.  From repeat measures of a handful of
stars, we found mean systematic offsets of $\sim$5-8 km s$^{-1}$
between observing runs.  These offsets were applied to all RVs from a
given run to place all measurements on the system of the Dec. 2007
velocities.

\section{ACS Candidate Selection}

The upper panel of Figure~\ref{fig:vhel_hist} shows measured
heliocentric radial velocities for all 224 stars in the SA 76 sample.
As already shown by \citetalias{gcm08}, the distribution has multiple
velocity peaks -- one broad and prominent peak at $\sim$30-40 km
s$^{-1}$ associated with Galactic stellar populations, and another at
$V_{\rm helio} \sim$ 90 km s$^{-1}$ identified as ACS debris.  In the
lower panel of Figure~\ref{fig:vhel_hist}, velocities are shown for
only faint ($g > 18.5$) stars, in order to focus on the MSTO region of
the ACS.  To facilitate comparisons to expected Galactic populations
in this work, we combined five realizations of the Besan\c{c}on Galaxy
model \citep{rrd+03} along this same line of sight.  Combining several
model queries serves to smooth over finite sampling statistics from
within each model population and reduce the noise in the predicted
distributions.  For comparison to our measured RVs, we overlay in
Figure~\ref{fig:vhel_hist} (as a filled histogram) an RV distribution
taken from the combined realizations of the Besan\c{c}on Galaxy model,
scaled to match the total number of stars observed.  The model
distribution was limited to the magnitude and color ranges of our
target selection (0.2$ < g-r < $0.6 and 18.5$ < g < $ 20.0) to sample
the same foreground/background populations (we note that this is the
same comparison done by \citetalias{gcm08}, but with additional
measured RVs).  The peak and dispersion of the observed RVs match the
model distribution fairly well, but with an additional peak
prominently visible at $V_{\rm helio} \sim $90 km~s$^{-1}$.  This peak
has already been identified by GCM08 (though using far fewer RV
measurements) as being due to ACS stream stars.

\begin{figure}%[!htp]
\plottwo{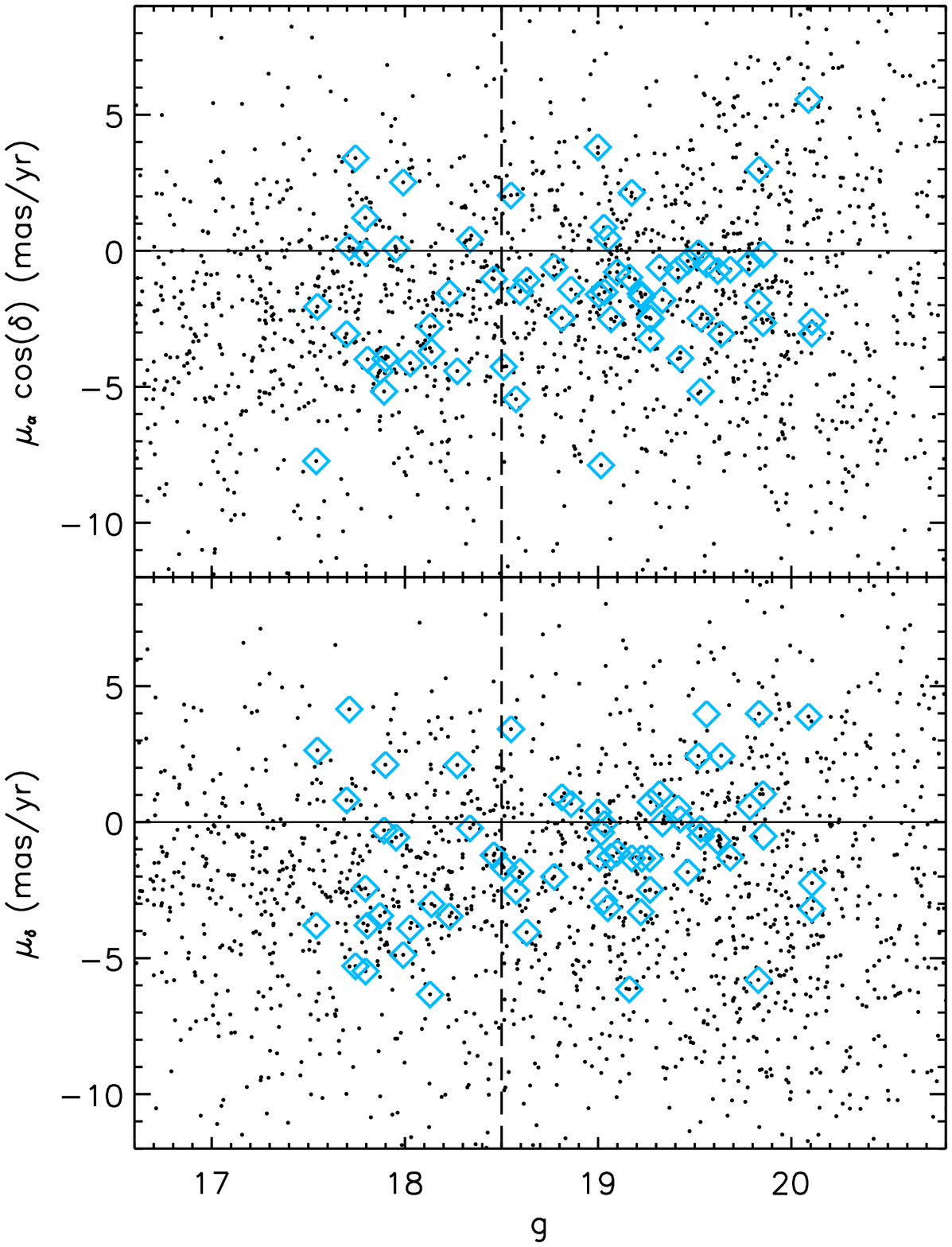}{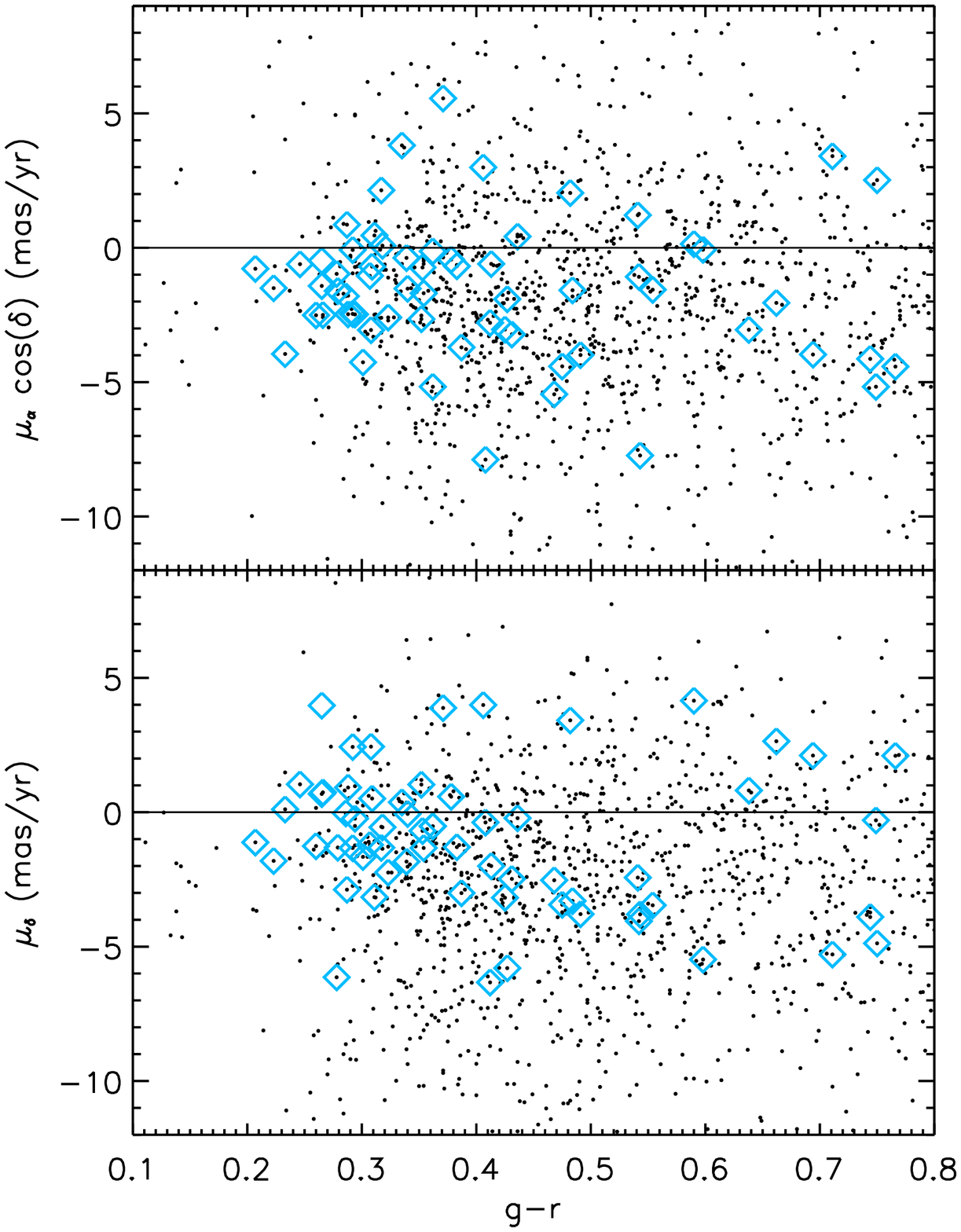}
\caption{Proper motions of all stars measured on at least four plates
(to remove spurious detections) as a function of $g$ magnitude (left)
and $g-r$ color (right).  Open (blue) diamonds are those stars within
our initial ACS candidate selection ($65 < V_{\rm helio} < 125$ km
s$^{-1}$).  Bright ($g <$ 18.5), predominantly red ($g-r \gtrsim 0.5$)
stars exhibit $\sim1.5\times$ (in RMS) larger scatter in both proper
motion dimensions than the fainter ACS main sequence candidates.
Because we are interested in the mean ACS motion, and not necessarily
identifying {\it all} ACS members in SA 76, we chose to remove bright
stars from the ACS candidates.  The vertical line at $g$ = 18.5 marks
the magnitude cut performed (in addition to the RPMD criteria
discussed in \S3) in selecting our final ACS
sample. \label{fig:mu_gmag_gr_targetselect}}
\end{figure}

ACS candidates are initially selected using all stars with 65$ <
V_{\rm helio} < $125 km~s$^{-1}$ (to ensure that all possible stream
members are included) -- a total of 87 stars.  Stars within this RV
selection are shown as solid triangles in the right-hand panels of
Figures~\ref{fig:cmd_targetselect} and \ref{fig:vpd_targetselect}.
Many of the stars thus selected are concentrated along the apparent
main sequence of the ACS population in the SDSS CMD, and are also more
tightly clumped in the VPD than the general Galactic populations.
Because the ACS velocity peak overlaps the Galactic distribution,
however, a selection of ACS candidates based solely on RVs will
contain some contamination from foreground (and background) Milky Way
stars.  Initially, we remove all stars having proper motions $|{\mu}|
\gtrsim 10$ mas~yr$^{-1}$ in either dimension.  Such proper motions
imply extremely large ($\gtrsim$ 500 km~s$^{-1}$) tangential
velocities if these stars are at distances $\sim$10 kpc.  Thus most of
the stars removed on this basis are nearby, high proper-motion Milky
Way disk stars.

\begin{figure}%[!htp]
\plotone{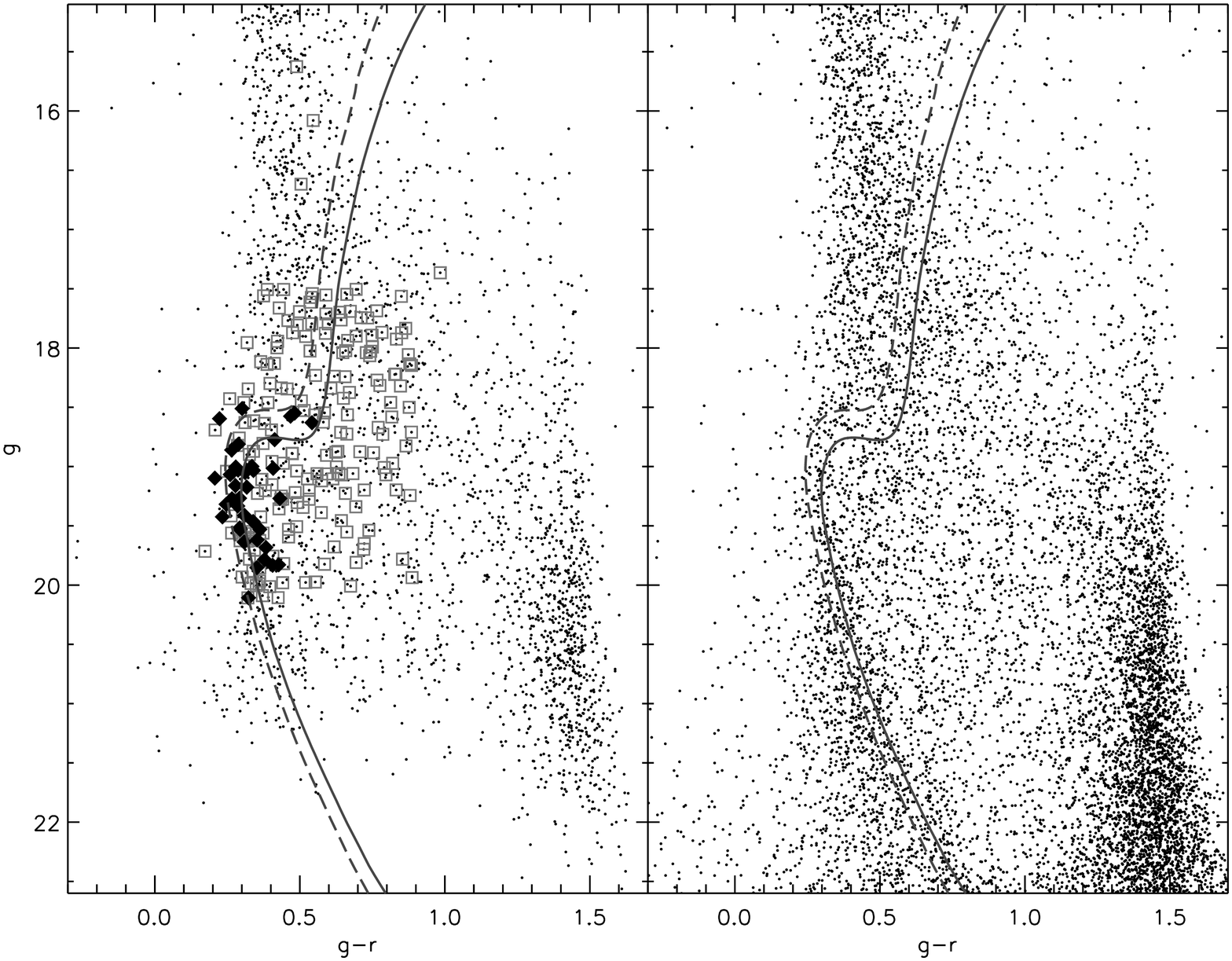}
\caption{{\it Left panel:} Color magnitude diagram of all stars in the
proper motion catalog (dots), with spectroscopic targets shown as open
squares.  The final ACS sample of 31 candidates is represented by
filled diamonds.  For comparison, two isochrones from \citet{dcj+08}
for old (10 Gyr) populations at a distance of 10 kpc are shown: the
solid line has [Fe/H] = -0.9, and the dashed line [Fe/H] = -1.3. Both
isochrones have been corrected for the mean reddening of $E(B-V) =
0.04$ from \citet{sfd98} for this line of sight.  {\it Right panel:}
All objects classified as stars in SDSS DR7, in a $1.3\arcdeg \times
1.3\arcdeg$ region centered on SA 76.  The ridgelines are the same as
in the left panel.  The main sequence of the Anticenter Stream can be
seen clearly among faint, blue stars.\label{fig:full_cmd}}
\end{figure}

\begin{figure}%[!htp]
\plotone{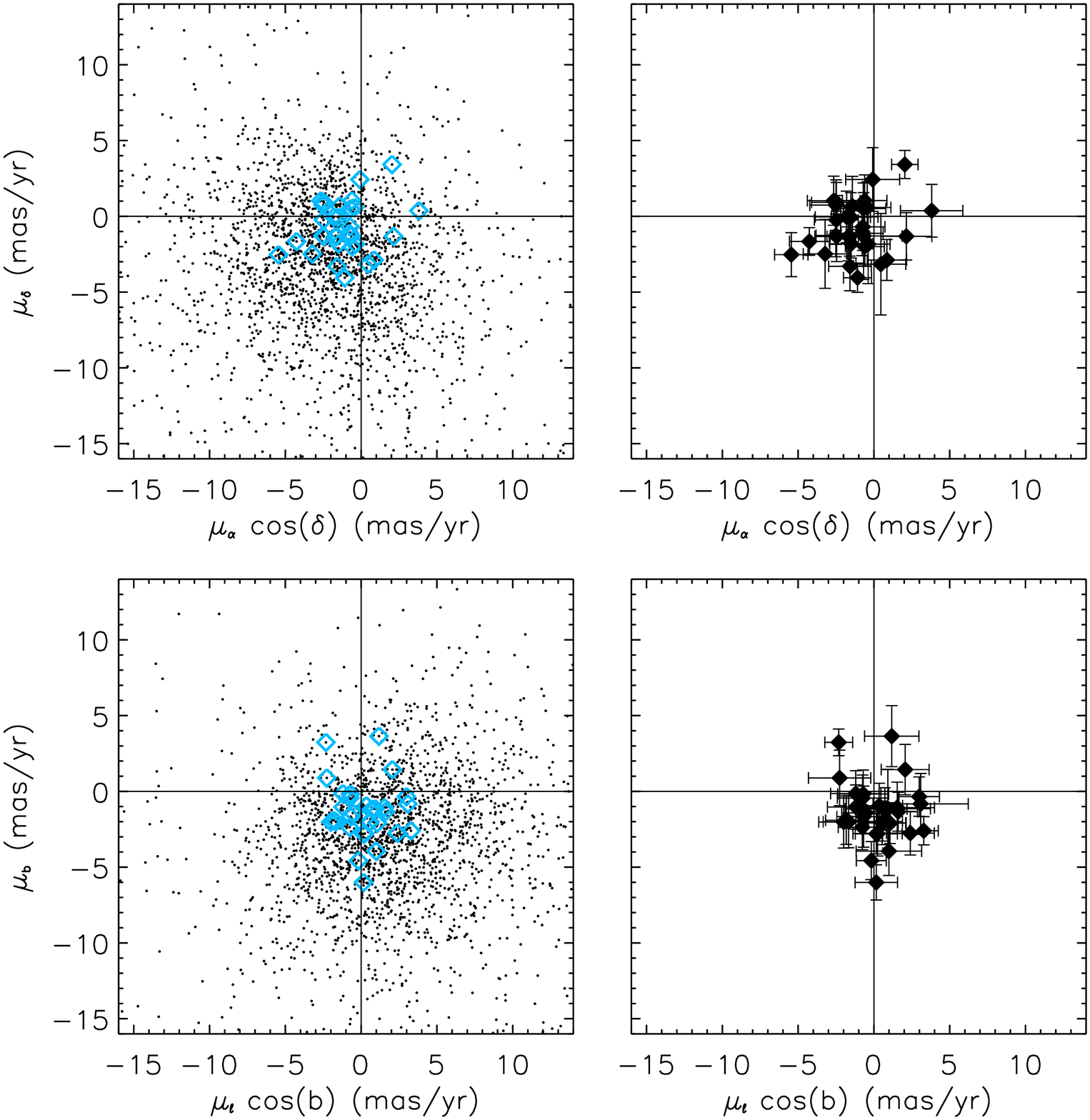}
\caption{{\it Left panel:} Vector point diagram of all well-measured
stars in SA 76 (black dots), with the ACS members overplotted as open
diamonds.  {\it Right panel:} Proper motions for only the ACS members,
with error bars reflecting individual formal
uncertainties. \label{fig:vpd_members}}
\end{figure}

Examination of the remaining sample of stars in a CMD (right panel of
Figure~\ref{fig:cmd_targetselect}) shows no obvious collection of
bright ($g < 18.5$) stars with a narrowly-defined RGB locus
representing an evolved counterpart to the well-defined MSTO.
Therefore, to distentangle co-moving stream members from
foreground/background Milky Way stars, we turn to the reduced proper
motion diagram (RPMD).  Reduced proper motion, first used extensively
by Luyten (e.g., \citealt{l22}), is defined as $H_m \equiv m + 5$ log
$\mu + 5$, where $m$ is apparent magnitude in a given bandpass (we
will use SDSS $g$ magnitudes hereafter), and $\mu$ is the total proper
motion in arcsec yr$^{-1}$.  Because the tangential velocity $V_{\rm
tan}$ (km~s$^{-1}) = 4.74*d*\mu$ (where $d$ is in pc and $\mu$ in
arcsec yr$^{-1}$), the reduced proper motion $H$ is analogous to an
absolute magnitude for stars of a common motion, and virtually
independent of distance.  Substituting terms and using the definition
of absolute $g$ magnitude, one finds that $H_g = M_g + 5$ log $V_{\rm
tan} - 3.38$.  Since we know the expected color-magnitude locus for a
given population, the reduced proper motion diagram ($H_g$ vs. $g -
r$) can be used to identify objects with similar tangential
velocities.  The RPMD for all objects measured on at least five plates
(to ensure well-measured proper motions) is shown in
Figure~\ref{fig:gr_rpmd}.  The left panel shows all well-measured
objects.  In the right panel, all spectroscopic targets are depicted
with open (grey) squares, with solid symbols representing stars within
the initial RV selection.  Stars in the RV-selected sample with red
(0.5 $< g-r <$ 0.9) colors and having reduced proper motion
inconsistent with that expected for the ACS main sequence and giant
branch were first removed from the sample; these are depicted as
filled green triangles in Figure~\ref{fig:gr_rpmd}.  It is apparent in
this plot that the remaining RV-selected candidates separate into two
groupings that resemble MSTO features at blue (0.2$ < g-r < $0.5)
colors at reduced proper motions of 11 $\lesssim H_g \lesssim$ 12.5
and 12.5 $\lesssim H_g \lesssim$ 14.5.  These groups were separated
into a ``lower'' (shown as solid blue squares in
Figures~\ref{fig:gr_rpmd} and \ref{fig:vpd_rpmdsel}) and an ``upper''
(filled red diamonds) RPMD sample, and examined to see whether there
are two populations at distinct $V_{\rm tan}$ within our RV selection.
The upper RPMD sample is tightly clumped in the proper motion vector
point diagram (Figure~\ref{fig:vpd_rpmdsel}, red filled squares),
while the lower RPMD sample shows considerable scatter inconsistent
with a common-motion stellar population at 10 kpc.  Because of the
wide dispersion in their proper motions, we surmise that the lower
RPMD sample is neither ACS debris nor any comoving, spatially
localized structure; we thus remove all of these stars from the sample
of ACS candidates.

Among this sample (red filled diamonds in
Figure~\ref{fig:vpd_rpmdsel}) there remain some outliers well outside
the proper motion clump.  These were examined to assess their
membership in the stream.  Proper motions in equatorial coordinates
are shown for the entire RV-selected sample as a function of $g$
magnitude and $g-r$ color in Figure~\ref{fig:mu_gmag_gr_targetselect}.
If the brightest, reddest stars in the initial sample selection are
red giants or subgiants associated with the ACS, they should have
tightly clumped proper motions at the same mean PM as the MSTO stars
(especially when one considers that the brighter stars should have
more precise proper motion measurements).  The contrary is true,
however -- bright, red stars exhibit much {\it more} scatter in both
proper motion dimensions than the tightly-clumped faint main sequence
stars.  Because this work is focused on measuring the mean motion of
the stream, and not necessarily identifying every possible member, and
because no obvious RGB population of the ACS is seen, all stars with
$g < 18.5$ were removed from the final set of PM candidates.  To
retain only well-measured stars, those stars measured on fewer than
five of the 15 plates were also culled from the sample, leaving a
total of 31 rather secure ACS members, based on all available
information.  These are confined to a thin, well-defined MSTO in the
CMD of Figure~\ref{fig:full_cmd} (left panel), which corresponds to
the obvious main sequence visible in the right panel.  This panel
shows all stars from SDSS DR7 in this field of view, with two
isochrones from \citet{dcj+08} for old (10 Gyr), metal-poor
([Fe/H]=-0.9 and [Fe/H]=-1.3; chosen to be close to the median
metallicities found for ACS members in \S4) populations at 10 kpc, the
distance we adopt for the ACS. The isochrones were also corrected for
the mean reddening $E(B-V) = 0.04$ \citep{sfd98} in the SA 76 field.
These ridgelines follow the clear overdensity of faint, blue stars, as
well as passing through the MSTO locus defined by the final sample of
identified ACS members.  The ACS members also clump tightly in the
proper motion VPD (Figure~\ref{fig:vpd_members}), as expected for a
distant, co-moving stellar population.

\begin{figure}%[!htp]
\epsscale{0.7}
\plotone{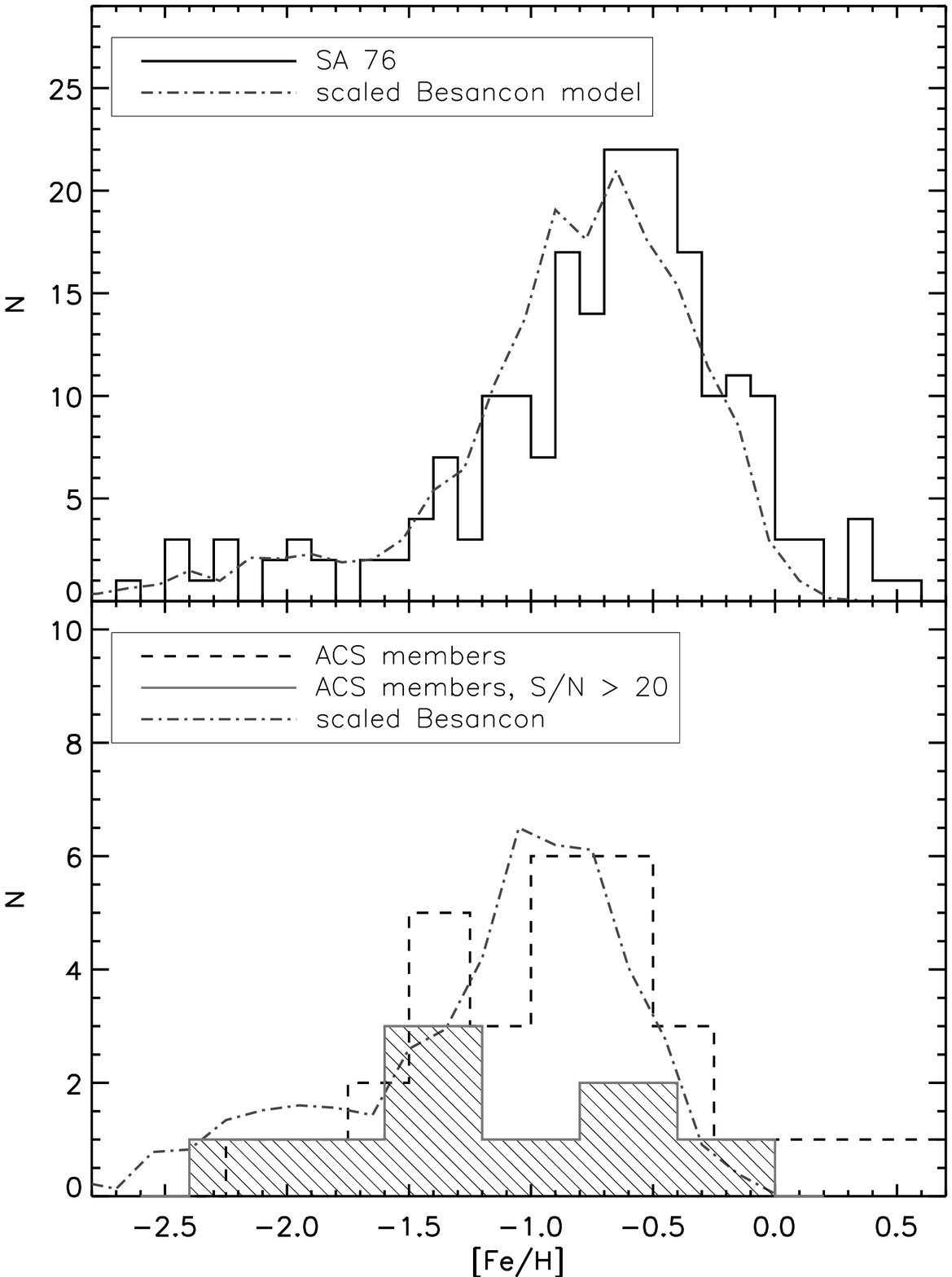}
\caption{{\it Upper panel:} Measured values of [Fe/H] for
spectroscopic targets observed in SA 76 (solid histogram).  The scaled
Besan\c{c}on model prediction for stars of similar magnitude and color
as our spectroscopic targets is shown as the dot-dashed line.  The
model reproduces the observed metallicity distribution quite well.
{\it Lower panel:} As in the upper panel, but with the dashed
histogram depicting only those stars in our final sample of ACS
members.  There is a hint of an excess of metal-poor
([Fe/H]$\lesssim$-1.3) stars in the overall ACS member sample compared
to model predictions, but we have too few ACS members to definitively
assess this possibility.  The solid, filled histogram shows the nine
stars with well-measured metallicity (i.e. $S/N >$ 20).  These are
concentrated more to the metal-poor end than the model distribution,
with a median [Fe/H] = -1.26, but with a broad abundance spread as
expected for, e.g., a dwarf galaxy stellar
population.\label{fig:feh_hist}}
\end{figure}

\section{Metallicities}

Stellar parameters are estimated for all spectroscopic targets using a
software pipeline developed by author JLC and others in the University
of Virginia stellar populations research group for this purpose.  This
pipeline uses fits to the distribution of [Fe/H] as a function of
low-resolution Lick Fe, Mg, and H$\beta$ indices from the library of
\citet[based on the spectra of \citealt{j98}]{s07} to determine
individual stellar metallicities.  Details about the code, which is
called EZ\_SPAM (Easy Stellar Parameters and Metallicities), will be
found in a forthcoming paper (Carlin et~al. 2010, {\it in prep.}).

Measured metallicities for all stars observed in SA 76 are shown in
the upper panel of Figure~\ref{fig:feh_hist}.  The expected
distribution (scaled to match the number of observed stars) from the
Besan\c{c}on galaxy model for stars with similar color and magnitude
distribution as our targets is shown for comparison (as we did for the
RV sample).  The distributions match quite well on visual inspection,
and a two-sided Kolmogorov-Smirnov (K-S) test finds a 79\% probability
that the two samples are drawn from the same parent population (we
note that this would likely be higher if a number of these stars were
not ACS members).  The lower panel of Figure~\ref{fig:feh_hist} shows
[Fe/H] for only the 31 stars identified as ACS members (dashed
histogram), again overlaying the scaled Besan\c{c}on distribution
(dot-dashed lines) for corresponding selection criteria.  The median
metallicity of the ACS members is [Fe/H] = -0.81, but with significant
scatter.  The median for the model is [Fe/H] = -0.91, and the K-S test
for these two samples suggests a 47\% probability that the ACS members
are drawn from the same population as the model distribution.
However, only 9 of these 31 spectra have $S/N >$ 20, so most of these
measurements of [Fe/H] should be regarded with some skepticism.  The 9
stars with well-measured metallicities (seen as a solid, filled
histogram in Fig.~\ref{fig:feh_hist}) yield median [Fe/H] = -1.26, and
include two stars with [Fe/H] $\lesssim$ -1.9.  Again, the spread in
metallicity is rather broad, as is seen in many dSph systems, where
there are often multiple epochs of star formation.

The apparent agreement between metallicities of the overall ACS member
sample with the Besan\c{c}on predictions (if real; see the caveat
regarding low $S/N$ of most of these spectra in the previous
paragraph) could arise for several reasons.  It may be that the ACS
``members'' are actually a sample of thick disk and/or halo stars
(noting that, at $b$=26.4$\arcdeg$, or $Z_{GC}=4.4$ kpc at a distance
of 10 kpc, near the Galactic anticenter, the line of sight toward SA
76 is not sampling many thin disk MSTO stars at faint magnitudes), and
thus agree with the [Fe/H] expected for smooth Galactic populations
because they are drawn from exactly those populations.  From the five
realizations of the Besan\c{c}on model used for comparison, we find
that a total of 214$\pm$13 stars are expected within the
color-magnitude selection initially applied (0.2$ < g-r < $0.6 and
18.5$ < g < $ 20.0).  Of these, only 4$\pm$2 are thin-disk stars, with
the rest made up of thick disk and halo populations.  In SA 76, we
find 271 stars within the same selection criteria (in a region of
equal area to that of the model), a significant ($\sim4\sigma$) excess
over the number of stars expected from the starcount models of smooth
Galactic populations that contribute to the Besan\c{c}on model.
Moreover, the ACS shows a well-defined narrow main sequence in the
SDSS CMD of Figure~\ref{fig:full_cmd} (or, more clearly, in Fig.~3 of
\citetalias{g06a}) indicative of a limited radial extent, unlike the
smoother and more broadly distributed stellar populations that may
arise if the ACS results from a warp or flare of the Galactic disk or
a flyby encounter with a massive perturber \citep{kbz+08,ybc+08}.
Furthermore, if the line of sight intersected a warped or flared MW
disk, it is unlikely that we would find the clear spatial separation
of the narrow ACS stream from the Galactic thin/thick disks seen in
\citetalias{g06a} (see also our Figure~\ref{fig:starcounts_orbit}
below).  It is thus unlikely that the ACS sample consists primarily of
the nominal MW stellar populations.  The second (and more likely)
possibility is that the ACS stars are metal-poor remnants of a
disrupted dSph, and thus resemble the metal-poor MW halo (thought to
be comprised mostly of stars from tidally disrupted dSphs and globular
clusters; e.g., \citealt{mmh94,bj05}) in the same magnitude and color
ranges.  Discrimination between these two scenarios could be provided
by high-resolution spectroscopic abundance analysis, wherein unique
abundance patterns may be able to distinguish differences between the
populations.

\section{Anticenter Stream Kinematics}
\subsection{Radial Velocities}

From the final selected ACS candidates, kinematical properties of the
ACS in SA 76 were estimated using a maximum likelihood method (e.g.,
\citealt{pm93,hgi+94,kwe+02}).  A systemic heliocentric radial
velocity of $V_{\rm helio} = 97.0\pm2.8$ km s$^{-1}$ was derived for
the ACS component, with an intrinsic dispersion $\sigma_{0} =
12.8\pm2.1$ km s$^{-1}$.  A Gaussian centered at this velocity, with
FWHM matching the measured dispersion, and encompassing the total
number of ACS candidates is overplotted as a dotted curve on the RV
histogram in the lower panel of Figure~\ref{fig:vhel_hist}.  The sum
of this Gaussian distribution and the scaled, binned Besan\c{c}on
model distribution (filled histogram) is plotted as the dot-dashed
histogram, which reproduces the observed velocities remarkably well.
This confirms our interpretation of the $\sim$97 km s$^{-1}$ peak,
which is not expected among smooth Galactic populations, as being due
to kinematically cold substructure.  The measured $V_{\rm helio}$
agrees with the measurement ($V_{\rm helio} = 88.8 \pm 5.0$ km
s$^{-1}$) of \citetalias{gcm08} at the 1.6$\sigma$ level (though we
remind the reader that those same data are also included in our sample
here), but we derive a dispersion more than twice the $\sigma_{0} =
5.9$ km s$^{-1}$ found by \citetalias{gcm08}.  This may mean that
\citetalias{gcm08} underestimated the intrinsic dispersion of the ACS
from their limited data set, or it may be that \citetalias{gcm08}
sampled one of the cold ``tributaries'' found by \citetalias{g06a} to
make up the larger Anticenter Stream.  In this latter scenario, the
superposition of multiple cold populations (whether from tributaries
or multiple orbital wraps) would lead to an overall larger measured
dispersion.  Whichever of these is the case, our measured $\sigma_{0}
= 12.8$ km s$^{-1}$ is typical of a tidal stream from a disrupted
dwarf galaxy (e.g., \citealt{mkl+04,mbb+07}); further high-resolution
study of large numbers of stars would be necessary to explore the
possibility of multiple stream tributaries within this field of view.
We also note that the $\sigma_{0} = 12.8$ km s$^{-1}$ dispersion we
measure in SA 76 is consistent with result $(\sigma_{0}\sim14.9$ km
s$^{-1}$) found by \citetalias{gcm08} for ACS debris $\sim23\arcdeg$
from SA 76 in their field "ACS-B" at $(\alpha, \delta)_{2000} \sim
(124\arcdeg, 37\arcdeg)$.

\subsection{Proper Motions}

\begin{deluxetable}{cccccccccc}%[!htp]
\tabletypesize{\scriptsize}
\tablecaption{Anticenter Stream Kinematics in SA 76\tablenotemark{a}\label{tab:kinematics}}
\tablewidth{0pt}
\tablehead{
\colhead{$V_{\rm helio}$} & \colhead{$V_{\rm GSR}$} & \colhead{$\sigma_0$} & \colhead{$\mu_{\alpha}$ cos($\delta$)\tablenotemark{b}} & \colhead{$\mu_{\delta}$} & \colhead{$\mu_{l}$ cos b} & \colhead{$\mu_{b}$} & \colhead{$U_{GSR}$\tablenotemark{c}} & \colhead{$V_{GSR}$} & \colhead{$W_{GSR}$} \\
\colhead{(km s$^{-1}$)} & \colhead{(km s$^{-1}$)} & \colhead{(km s$^{-1}$)} & \colhead{(mas yr$^{-1}$)} & \colhead{(mas yr$^{-1}$)} & \colhead{(mas yr$^{-1}$)} & \colhead{(mas yr$^{-1}$)} & \colhead{(km s$^{-1}$)} & \colhead{(km s$^{-1}$)} & \colhead{(km s$^{-1}$)}
}
\startdata
 97.0$\pm$2.8 & -6.3$\pm$2.8 & 12.8$\pm$2.1 & -1.20$\pm$0.34 & -0.78$\pm$0.36 & 0.23$\pm$0.36 & -1.43$\pm$0.34 & -86.7$\pm$12.2 & 158.5$\pm$17.0 & -10.5$\pm$23.4 \\
\enddata
\tablecomments{All calculations assume d$_{ACS}$ = 10.0 kpc, and $V_{circ}$ = 220 km s$^{-1}$ at R$_{0}$ = 8.5 kpc. This yields right-handed, Galactocentric coordinates for this field of $(X, Y, Z)_{GC}$ = (-16.32, -4.38, 4.44) kpc.}
\tablenotetext{a}{($\alpha$, $\delta$)$_{2000}$ = (125.28$\arcdeg$,14.69$\arcdeg$); (l, b) = (209.25$\arcdeg$,26.36$\arcdeg$)}
\tablenotetext{b}{Errors in individual stellar proper motions contain $\sim$0.4 mas/yr zero point uncertainty added in quadrature to the uncertainty in the measured PM.}
\tablenotetext{c}{We used the Dehnen \& Binney 1998 values for the solar peculiar motion: ($U_0$, $V_0$, $W_0$) = (10.00, 5.25, 7.17)$\pm$(0.36,0.62,0.38) km s$^{-1}$ (in a right-handed frame).}
\end{deluxetable}

From this same sample of 31 ACS candidates, maximum likelihood
estimates were derived for the absolute proper motions of
$\mu_{\alpha}$ cos $\delta = -1.20\pm0.34$ mas yr$^{-1}$, and
$\mu_{\delta} = -0.78\pm0.36$ mas yr$^{-1}$.  The uncertainties of
individual stars used in these estimates contain the error in the
proper motion zero point (derived from galaxies and QSOs) added in
quadrature to the measurement uncertainty of each individual relative
PM.  Both of these PM measurements differ significantly from those of
GCM08; our $\mu_{\alpha}$ cos $\delta$ is lower by 1.87 mas yr$^{-1}$,
and $\mu_{\delta}$ lower by 1.51 mas yr$^{-1}$.  The GCM08 estimate
was based on 16 RV members, with proper motions from SDSS/USNO-B
\citep{mml+08,mml+04}.  The SDSS/USNO-B PMs have individual
uncertainties of $\sim$4 mas yr$^{-1}$, so we regard the GCM08
estimates as only rough limits on the tangential motions of the
stream.  Our current absolute PMs are derived from much more precise
(1-2 mas yr$^{-1}$ per star) PMs, as well as a larger number of
securely identified members; we thus consider these new estimates more
reliable.  We note, however, that although the components of the
implied space velocities from our measurements differ significantly
from those of \citetalias{gcm08} (mostly in the component vertical to
the disk), the magnitude of the motion differs by only $\sim$40 km
s$^{-1}$ (assuming a 10 kpc distance).  Converted to PMs along
Galactic coordinates, our results become $\mu_{l}$ cos~$b =
0.23\pm0.36$ mas yr$^{-1}$, and $\mu_{b} = -1.43\pm0.34$ mas
yr$^{-1}$.

\begin{figure}%[!htp]
\epsscale{1.0}
\plotone{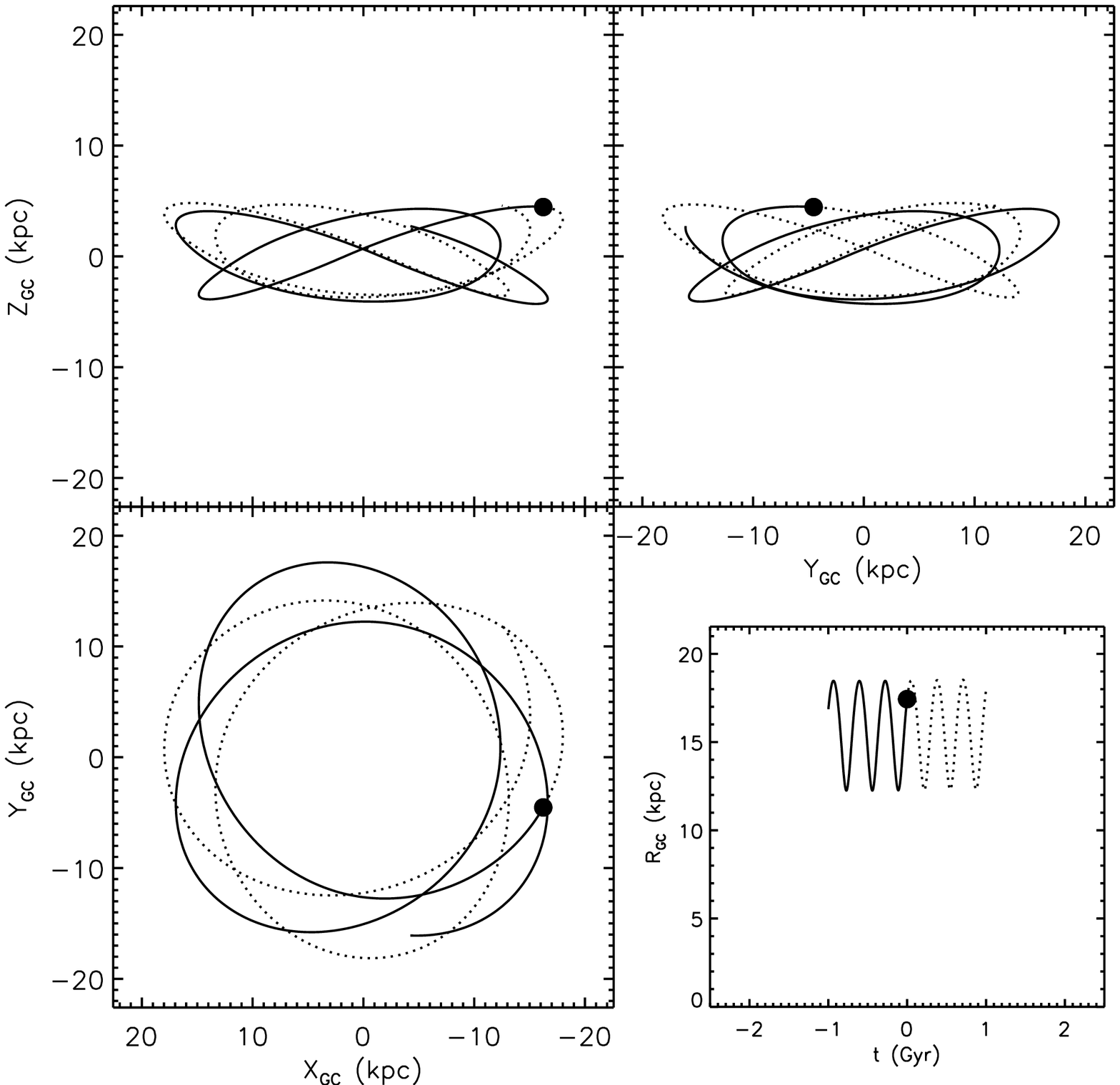}
\caption{Orbit based on the measured ($U,V,W$) space velocity in SA
76.  The distance was taken to be 10 kpc at the position of SA 76, and
the orbit integrated both backward (solid lines) and forward (dotted
curves) for 1 Gyr (about 2.5 orbits) to trace out the full orbital
path.\label{fig:orbit_xyz}}
\end{figure}

\subsection{Three Dimensional Motion}% and Orbit}

The measurements of all three components of stream star motions in SA
76 were used to estimate the orbital characteristics of the Anticenter
Stream.  Some information about the stream distance is needed to
convert position on the sky to a Galactocentric position.
\citet{g06a} estimated the distance to the ACS to be 8.9$\pm$0.2 kpc,
based on the required shift of an M13 fiducial to match the ACS main
sequence.  Though this provided a peak in the matched filtering, the
metallicity of M13 ([Fe/H] = -1.54; \citealt{h96}) is lower than the
median we measured of [Fe/H] $\approx$ -0.8 for ACS members ([Fe/H]
$\approx$ -1.3 for the 9 high $S/N$ members; see \S 4 for details).
We suggest that a slightly more metal-rich isochrone than that of M13
would be a more reasonable fit.  This would place the stream stars
slightly farther away than the $\sim$9 kpc from \citetalias{g06a}.
Given the uncertainty in the stream distance, we adopt 10$\pm$3 kpc
for the distance to the ACS in SA 76 in all further calculations in
this work.

The prescription of \citet{js87} was used to derive the Cartesian
Galactic space velocity components ($U$, $V$, and $W$) in a
right-handed frame with the origin at the Galactic center (i.e. $U$ is
positive toward the Galactic center, $V$ in the direction of Galactic
rotation, and $W$ toward the North Galactic Pole).  The updated
transformation matrix for J2000 coordinates was taken from the
Hipparcos catalog introduction, and the correction to the Local
Standard of Rest was done by removing the solar motion of
\citet{db98}: ($U$,$V$,$W$)$_{Sun}$ = (10.00, +5.25, +7.17)$\pm$(0.36,
0.62, 0.38) km s$^{-1}$.  A circular velocity of $V_{circ}$ = 220 km
s$^{-1}$ was added to the derived velocities to convert from the
heliocentric frame to Galactocentric frame velocities of
($U$,$V$,$W$)$_{GSR}$ = (-86.7, 158.5, -10.5)$\pm$(12.2, 17.0, 23.4)
km s$^{-1}$, where the uncertainties were derived from the errors in
the three velocity measurements, again following the Johnson \&
Soderblom method (see Table~\ref{tab:kinematics} for a summary of
these kinematical quantities for ACS members in SA 76).

\section{Orbit From ACS Candidates: Why Isn't the Motion Aligned With the Visible Stream?}

The orbit for the ACS candidates in SA 76 was integrated in the
\citet{jsh95} Galactic potential, which consists of a \citet{h90}
spheroidal bulge, a disk following the \citet{mn75} prescription, and
a logarithmic halo.  As for all calculations in this work, we adopt
$R_{\sun}$ = 8.5 kpc and $V_{circ}$ = 220 km s$^{-1}$.  Orbit
derivation was started at the position of SA 76 with the measured
($U$,$V$,$W$)$_{GSR}$ components and a distance of 10 kpc, and was
integrated both forward and backward for 1 Gyr, in order to map out
the entire path and kinematics of the stream orbit.  The orbit is
shown in Galactic Cartesian ($X$,$Y$,$Z$)$_{GC}$ coordinates in
Figure~\ref{fig:orbit_xyz}, with the large dot representing the
assumed position of ACS stars in SA 76, the solid line the backward
orbit integration, and the dotted line the forward integration.  This
orbit has peri- and apo-galactic radii of $R_{p}$ =
12.3$^{+2.3}_{-2.2}$ kpc and $R_{a}$ = 18.5$^{+1.2}_{-0.4}$ kpc, and
an eccentricity of $e \equiv (R_{a} - R_{p})/(R_{a} + R_{p}) =
0.20^{+0.08}_{-0.05}$, where errors have been calculated by comparing
orbits with the maximum and minimum space velocities from
uncertainties in $U$, $V$, and $W$.  Taking our measured $\vec{R}$ =
($X$,$Y$,$Z$) and $\vec{V_{tot}}$ = ($U$,$V$,$W$), we can derive the
angular momentum (per unit mass) for this mean motion, $\vec{L} =
\vec{R} \times \vec{V_{tot}}$.  This yields a $Z$-component of angular
momentum per unit mass of $L_Z \sim -2960\pm330$ kpc km s$^{-1}$,
placing the SA 76 ACS debris on a prograde orbit ($L_Z > 0 \equiv$
retrograde, $L_Z < 0 \equiv$ prograde).  The inclination of the orbit
is estimated here as cos $i$ = -$L_Z/L$, resulting in $i=
16.2^{+3.4}_{-1.4}$ degrees.

\begin{figure}%[!htp]
\plottwo{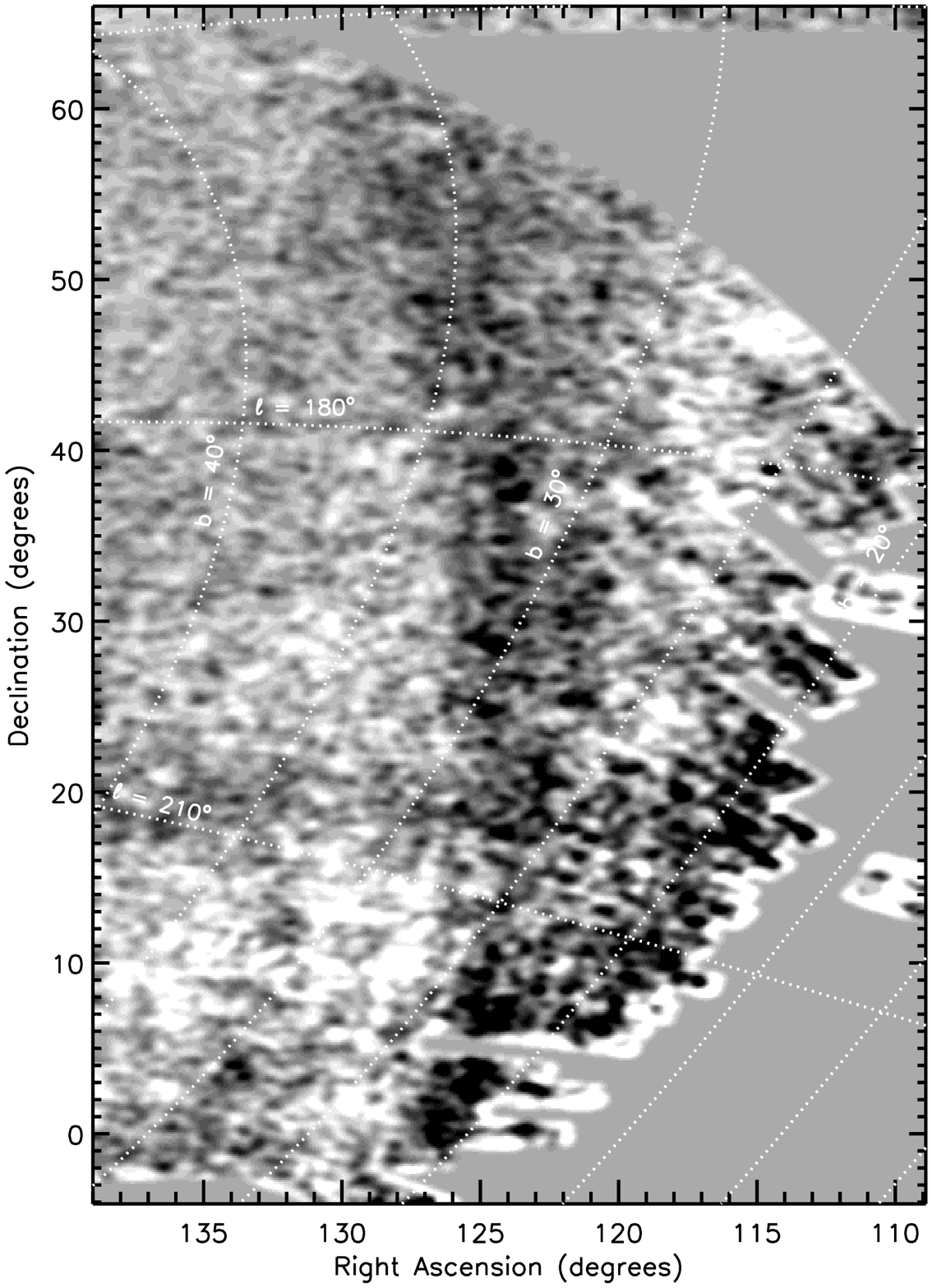}{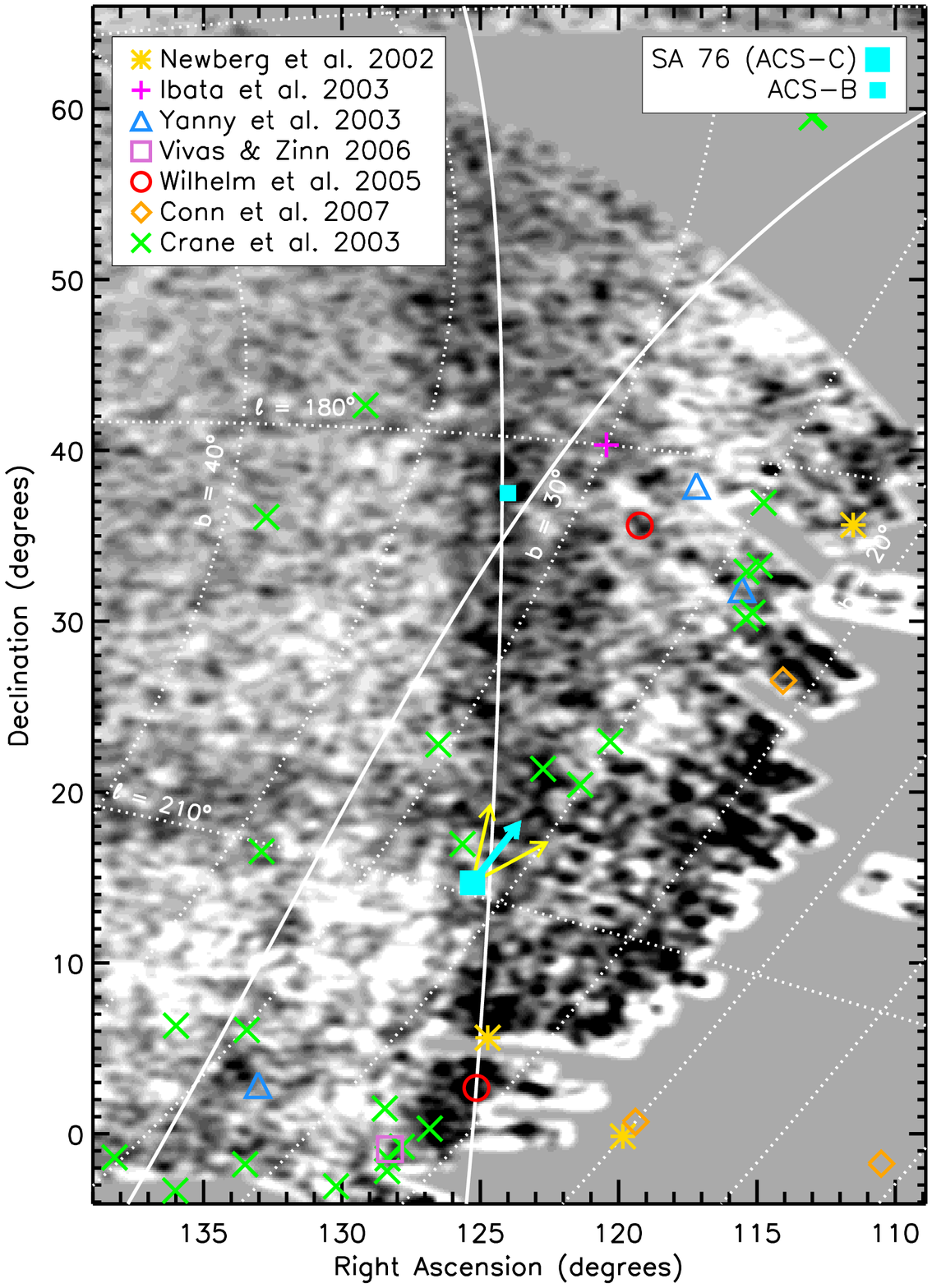}
\caption{Map of filtered star counts in the western portion of the
SDSS footprint, showing the prominent Anticenter Stream spanning the
entire vertical extent of the SDSS database in this region.  This plot
is similar to that of \citetalias{g06a}, but here using SDSS DR7.  In
the right-hand panel, the best-fitting orbit from \citetalias{gcm08}
(solid line) is overlaid; this orbit was constrained to fit 30
positions along the vertical stream, as well as radial velocities in
the two fields denoted by filled squares.  The large (cyan) square at
$(\alpha, \delta)\approx(125.3\arcdeg,14.7\arcdeg$) represents SA 76,
the field studied in this paper, with the large (cyan) arrow a vector
corresponding to the measured proper motions in that field.  The
motion implied by the 3$\sigma$ uncertainties in ACS kinematics are
represented by the yellow vectors flanking the mean motion vector.
The kinematics of SA 76 ACS debris apparently trace back in the
direction of the EBS (the feature at $[\alpha, \delta] \sim
[134\arcdeg,3\arcdeg]$), though no continuous overdensity is seen
connecting the EBS to the ACS.  As discussed in Section 6.5 and
illustrated in Figure~\ref{fig:starcounts_weststream}, the (cyan)
vector representing our measured motion, while misaligned with the
\citetalias{gcm08} orbit, seems to follow the western tributary of the
ACS pointed out by \citetalias{g06a}.  Colored symbols show the many
purported detections of Monoceros debris in this region of the sky,
most of which are not spatially coincident with the ACS.
\label{fig:starcounts_orbit}}
\end{figure}

To convert to a Galactocentric rest frame, we must remove the
contribution of the Sun's 220 km s$^{-1}$ circular velocity ---
($\mu_{\alpha}$ cos $\delta$, $\mu_{\delta}) = (0.45, -4.20)$ mas
yr$^{-1}$ for a non-moving object at 10 kpc along the SA 76 line of
sight --- to the measured proper motions. By subtracting this from the
mean measured proper motion of ACS candidates in SA 76, we derive a
Galactocentric proper motion of ($\mu_{\alpha}$ cos $\delta$,
$\mu_{\delta})' = (-1.65, 3.42) \pm (0.34,0.36)$ mas yr$^{-1}$.
The ACS is a distinct stream running almost North-South (in celestial
coordinates) on the sky, so it is curious that the proper motion we
derive from stars selected to be members of this stream is not oriented
along that same N-S direction.  This is evident in
Figure~\ref{fig:starcounts_orbit}, which shows the best-fit orbit from
\citet[solid line]{gcm08} overlying an image of the matched-filter
starcount density similar to that of \citet{g06a}, but using SDSS DR7.
The \citetalias{gcm08} orbit was constrained by positions and radial
velocities in two fields (ACS-B and ACS-C, shown as the small and
large filled cyan squares in the figure), as well as 30 positions
along the visible portion of the stream.  This orbit traverses the
vertical (i.e. roughly constant right ascension) stream, and passes
through the overdensity between $\alpha \approx$ 132-137$\arcdeg$,
$\delta \approx$ 0-10$\arcdeg$ known as the Eastern Banded Structure
(EBS; \citetalias{g06a}) on the subsequent orbital wrap.  The large
(cyan) arrow centered on the SA 76 position is a vector representing
the magnitude and direction of the mean (Galactocentric) proper
motions measured in SA 76, with yellow arrows on either side
representing the 3$\sigma$ uncertainty in the total space motion.  The
motion is not aligned along the obvious stream (though it is
consistent with being so at the $\sim3\sigma$ level), but traces a
path roughly along the orientation from the EBS to SA 76.  In this
section, we will explore in detail some possible explanations for the
misalignment of our proper motion and the visible path of the
Anticenter Stream.

\subsection{Comparison to \citet{gcm08} Orbit For the ACS}

The orbit we have derived from kinematics of ACS candidates has peri-
and apo-galactic radii of $R_{p}$ = 12.3$^{+2.3}_{-2.2}$ kpc and
$R_{a}$ = 18.5$^{+1.2}_{-0.4}$ kpc, and an eccentricity of $e =
0.20^{+0.08}_{-0.05}$.  This is a slightly more elongated orbit than
the one found by \citetalias{gcm08} for the ACS, which was found to
have $e \sim 0.1$, but with a nearly identical apogalacticon (18.5 kpc
vs. 19.0$\pm$1.9 kpc from \citetalias{gcm08}).  \citetalias{gcm08}
ruled out a retrograde orbit for the ACS based on SDSS/USNO-B proper
motions; we further note that the retrograde model in that work
predicted extremely large magnitude proper motions ($\mu_{\delta} \sim
-10$ mas yr$^{-1}$) in ACS-C (SA 76) that we can definitively rule out
based on our current measurements.  The inclination of our orbit for
ACS candidates was found to be $i=16.2^{+3.4}_{-1.4}$ degrees.  This
inclination angle is seemingly consistent with the result ($i =
20.1\arcdeg$) from \citetalias{gcm08}; however, we note that the
estimates derived in this way do not give the orientation of the
orbital plane, but only the angle it makes to the $XY$ plane.  At the
position of SA 76, we find ($U,V,W)_{\rm GSR}$ = (-52.4, 207.4, 79.0)
km s$^{-1}$ for the \citetalias{gcm08} orbit (based on kinematical
quantities from their Table 2).  The relative $Z$-components of the
total velocity, $W_{GSR}$, for this orbit and our proper
motion-derived result are in opposite directions, indicating that the
\citetalias{gcm08} orbit is oriented away from the Galactic disk at $i
\sim 20\arcdeg$, while the proper motions indicate movement downward
toward the disk at $i \sim 20\arcdeg$ (this difference in orientation
can be seen in Figure~\ref{fig:starcounts_orbit}).

\citetalias{gcm08} found that a stream orbit constrained by thirty
fiducial points along the spatial distribution and their measured RVs
in two fields along the stream should produce proper motions at the
position of ACS-C (SA 76) of ($\mu_{\alpha}$ cos $\delta$,
$\mu_{\delta}$) $\sim$ (0.64, 0.67)$\pm$(0.03, 0.35) mas yr$^{-1}$
(for the model unconstrained by SDSS/USNO-B proper motions).  This
disagrees with our measurements by (1.84, 1.45) mas yr$^{-1}$, or
$\sim5\sigma$ in $\mu_{\alpha}$ cos $\delta$ and $\sim4\sigma$ in
$\mu_{\delta}$.  Such a discrepancy cannot be due to the $\sim$0.43
mas yr$^{-1}$ uncertainty in our SA 76 proper motion zero point, which
is robustly determined by $\gtrsim$100 point-like galaxies in each
field.  An offset of this magnitude ($\Delta \mu \approx$ 2.4 mas
yr$^{-1}$ in total PM) corresponds to a difference in tangential
velocity of $>$110 km s$^{-1}$ at 10 kpc, so it is unlikely to be
simply a measurement error.  As will be discussed in \S6.2, comparison
of our measurements with the expected PMs along this line of sight
from the Besan\c{c}on model shows no obvious offset between the
model-predicted motions of smooth Galactic populations and our
results, so we do not believe a systematic proper motion error is
present in our derivations.  However, the total 3-D velocity we derive
differs by only $\sim$47 km s$^{-1}$ ($\sim1.5\sigma$ using the uncertainty in our measured $UVW$ velocity) from the \citetalias{gcm08} orbit
at the position of SA 76, and we will argue further in \S6.5 that the
two results are consistent with having originated from separate
components of a substructured progenitor (e.g., a dwarf galaxy with
associated globular clusters), but that in SA 76 we may not be
measuring the motion of the main body of the stream.

\subsection{Is the "Misalignment" Due to Systematic Errors in the Proper Motions?}

SA 76 lies along the visible stream (though apparently on the
periphery of the ACS rather than on a region of highest density; see
Figures~\ref{fig:starcounts_orbit} and
\ref{fig:starcounts_weststream}), and inspection of the
color-magnitude diagram shows a clear overdensity of faint, blue stars
in this field.  Stream candidates were selected from our SA 76
proper-motion database to be consistent with membership in a structure
at $\sim$10 kpc distance.  We have shown (\S~3) that the expected
velocity distribution of field stars (including both MW halo and thick
disk) among our selection criteria is reproduced well, so that the
significant number of ACS candidates in the narrow velocity peak
suggests that the overdensity is real.  The clear narrow peak among
radial velocities of these candidates confirms that we are indeed
sampling members of a co-moving population among these candidates.
Because the stream traces such a distinct swath across the sky, we
expect the tangential motions of stars within the identified velocity
peak in SA 76 to follow this stream.  The fact that the 3-D space
velocity of the ACS candidates is directed at a $\sim$30-degree angle
to the visible stream led us to explore the possibility that there was
a systematic error in our proper motion measurements.  

\begin{figure}%[!htp]
\plotone{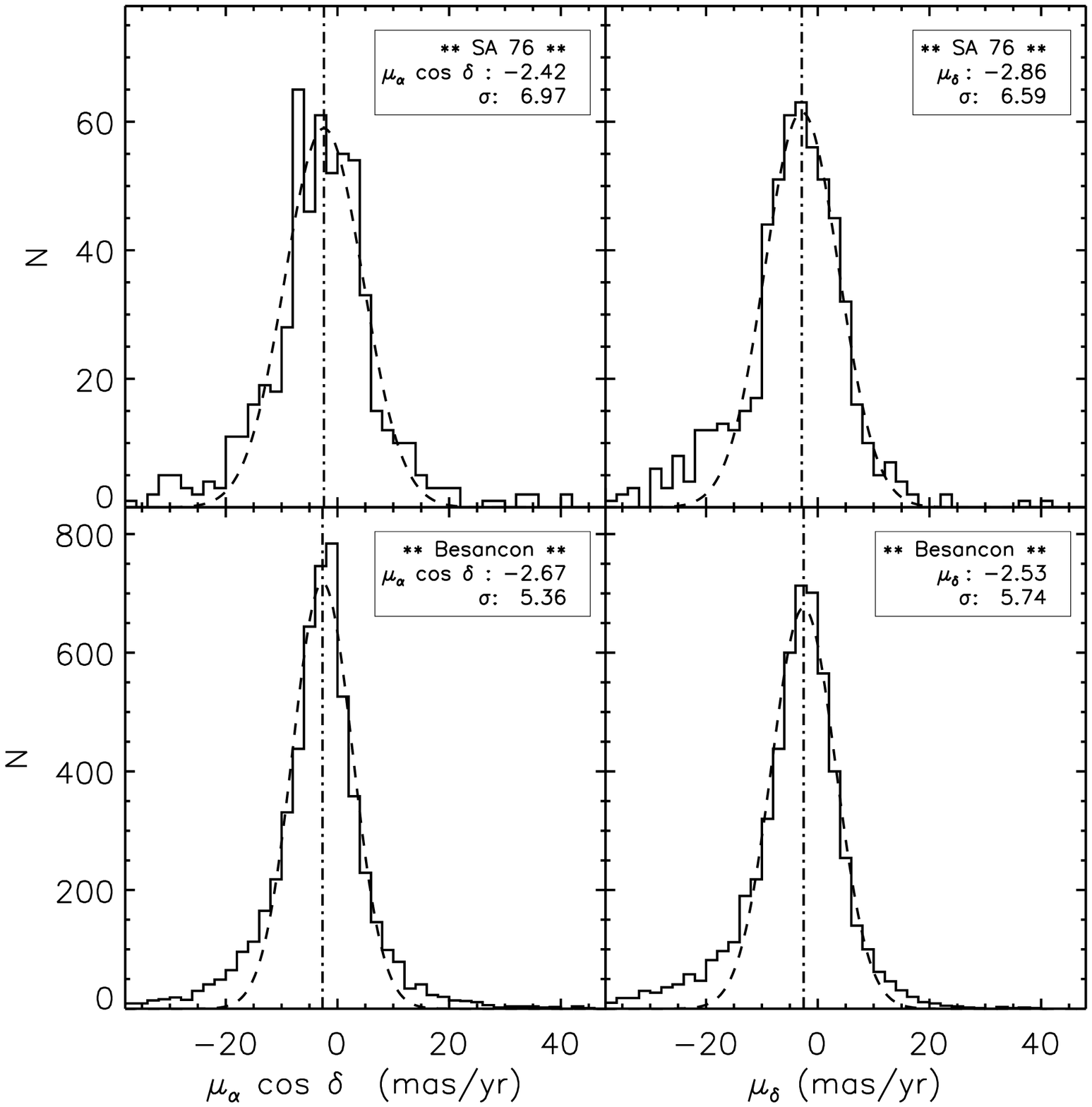}
\caption{Proper motions in each dimension for faint ($18 < g < 21$),
red ($1.1 < g - r < 1.7$) stars (selected to be nearby M dwarfs) in SA
76 (upper) and Besancon (lower).  The agreement between the fitted
mean values for $\mu_{\alpha}$ cos $\delta$ and $\mu_{\delta}$ for the
two samples leads us to conclude that no large systematic error is
present in our measured proper motions. \label{fig:sa76_besancon_pms}}
\end{figure}

\subsubsection{Comparison to the Besan\c{c}on Galaxy model}

Because we have reliably measured proper motions for most $g \lesssim
20.5$ stars in the field of view, and many as faint as $g = 21$, we
chose to use stars well outside the ACS candidate selection regions to
assess how well we recover the expected motions of Galactic stellar
populations.  To do this, we selected faint ($18 < g < 21$), red ($1.1
< g - r < 1.7$) stars from both the SA 76 proper motion catalog and
the Besan\c{c}on model predictions for the same line of sight.  At
these red colors, nearly all the faint stars thus selected should be
nearby (foreground) Milky Way M-dwarfs (and contain no Anticenter
Stream debris).  Figure~\ref{fig:sa76_besancon_pms} shows the
equatorial proper motions in each dimension for these M-dwarf
candidates -- the upper panels depict measurements in SA 76, and the
lower panels the model predictions.  Gaussian fits to these
distributions (dashed lines in the figure) have peaks at
($\mu_{\alpha}$ cos $\delta,
\mu_{\delta})_{\rm SA 76}$ = (-2.42, -2.86)$\pm$(0.27, 0.25) mas
yr$^{-1}$ and ($\mu_{\alpha}$ cos $\delta, \mu_{\delta})_{\rm model}$
= (-2.67, -2.53)$\pm$(0.16, 0.17) mas yr$^{-1}$.  These agree at the
$\sim1\sigma$ level, and differ by less than the $\sim$0.4 mas
yr$^{-1}$ uncertainty in our proper motion zero point.  We further
examined our proper motions for residual color- and
magnitude-dependent systematics, and found no significant trends (as
evidenced by the lack of slope in the mean PMs with either magnitude
or color in Figure~\ref{fig:mu_gmag_gr_targetselect}).

To investigate whether contaminant Milky Way thick disk and halo stars
are skewing our mean proper motions for ACS candidates, we selected
all stars from the Besan\c{c}on model within the same color and
magnitude criteria as our ACS candidates ($18.5 < g < 20.5$, $0.2 < g
- r < 0.6$).  Gaussian fits to the proper motions of these selected
model populations yield ($\mu_{\alpha}$ cos $\delta,\mu_{\delta})_{\rm
model}$ = (-0.36, -2.66)$\pm$(0.06, 0.09) mas yr$^{-1}$.  This differs
by $\sim(2.5,5.2)-\sigma$ from our mean ACS proper motions of
($\mu_{\alpha}$ cos $\delta, \mu_{\delta})_{\rm SA 76}$ = (-1.20,
-0.78)$\pm$(0.34, 0.36) mas yr$^{-1}$. These discrepancies are much
larger than any expected due to uncertainties in the proper motion
zero point, and, as shown in the previous paragraph, are unlikely to
result from a systematic offset in the proper motions.  Thus, any
non-stream members (i.e. thick disk or halo stars) that have been
misidentified as ACS candidates have likely skewed the proper motion
measurements to higher values in $\mu_{\alpha}$ cos $\delta$, and
shifted our estimate of $\mu_{\delta}$ lower than its intrinsic value.
The component of proper motion along right ascension one would measure
for an object at a distance of 10 kpc moving along the visible stream
(i.e. with motion along constant right ascension) would be
$\mu_{\alpha}$ cos $\delta \approx$ 0.68 mas yr$^{-1}$, which differs
by $\sim5.5\sigma$ from our measured value for ACS candidates.  Our
measurement for candidate ACS debris stars is lower than both this
prediction and the expected proper motions of Milky Way populations
along this line of sight.  Thus the misalignment of our measured PM
from the expected stream motion cannot be due to contamination of our
sample by foreground Milky Way stars, which would actually bring our
measurements {\it closer} to the expected stream motion.  In addition,
the magnitude of the Besan\c{c}on-predicted total proper motion,
$\mu_{\rm model}$ = 2.68 mas yr$^{-1}$, is nearly twice that of the
proper motion we measured ($\mu_{\rm SA 76}$ = 1.43 mas yr$^{-1}$) for
ACS debris -- a difference of $\sim3.5\sigma$.  The difference between
our mean proper motions and the expected stream and foreground star
kinematics is significantly larger than can be accounted for by the
uncertainties in the proper motion zero point, which are at the
$\sim$0.4 mas yr$^{-1}$ level in each PM dimension.  Based on all of
the above arguments, we conclude that the difference between the
orientation of the 3-D space motion we measure for purported ACS
debris and the expected direction of motion is not due to systematic
errors in the proper motions, but is a real kinematical difference.

\begin{figure}%[!htp]
\epsscale{0.85}
%\plotone{acsmembers_sa76_vs_sdss_vpd_noerrorbars.gx.crop.eps}
%\plotone{acsmembers_sa76_vs_sdss_vpd_noerrorbars.bluestars.crop.eps}
\plotone{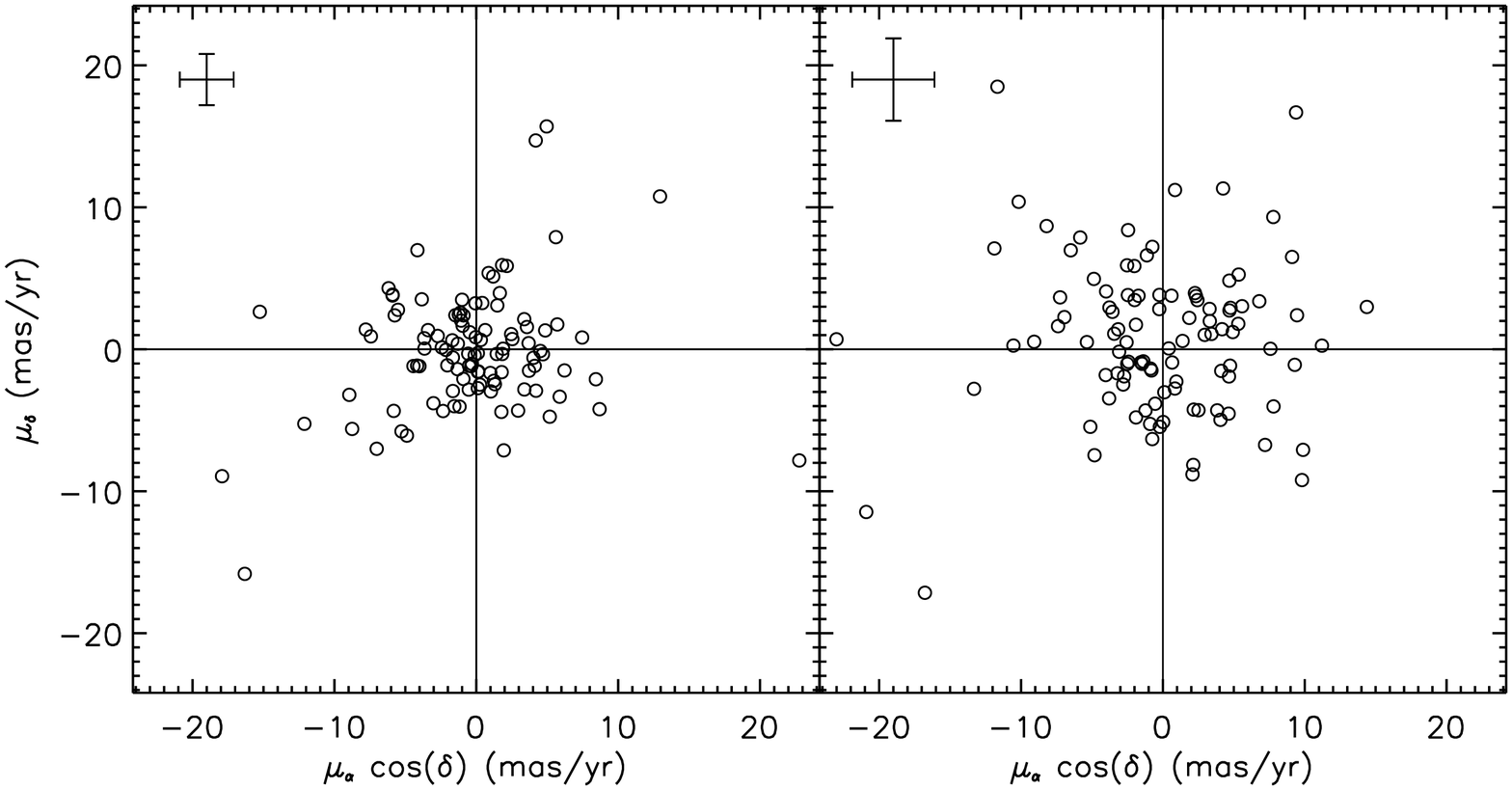}
\plotone{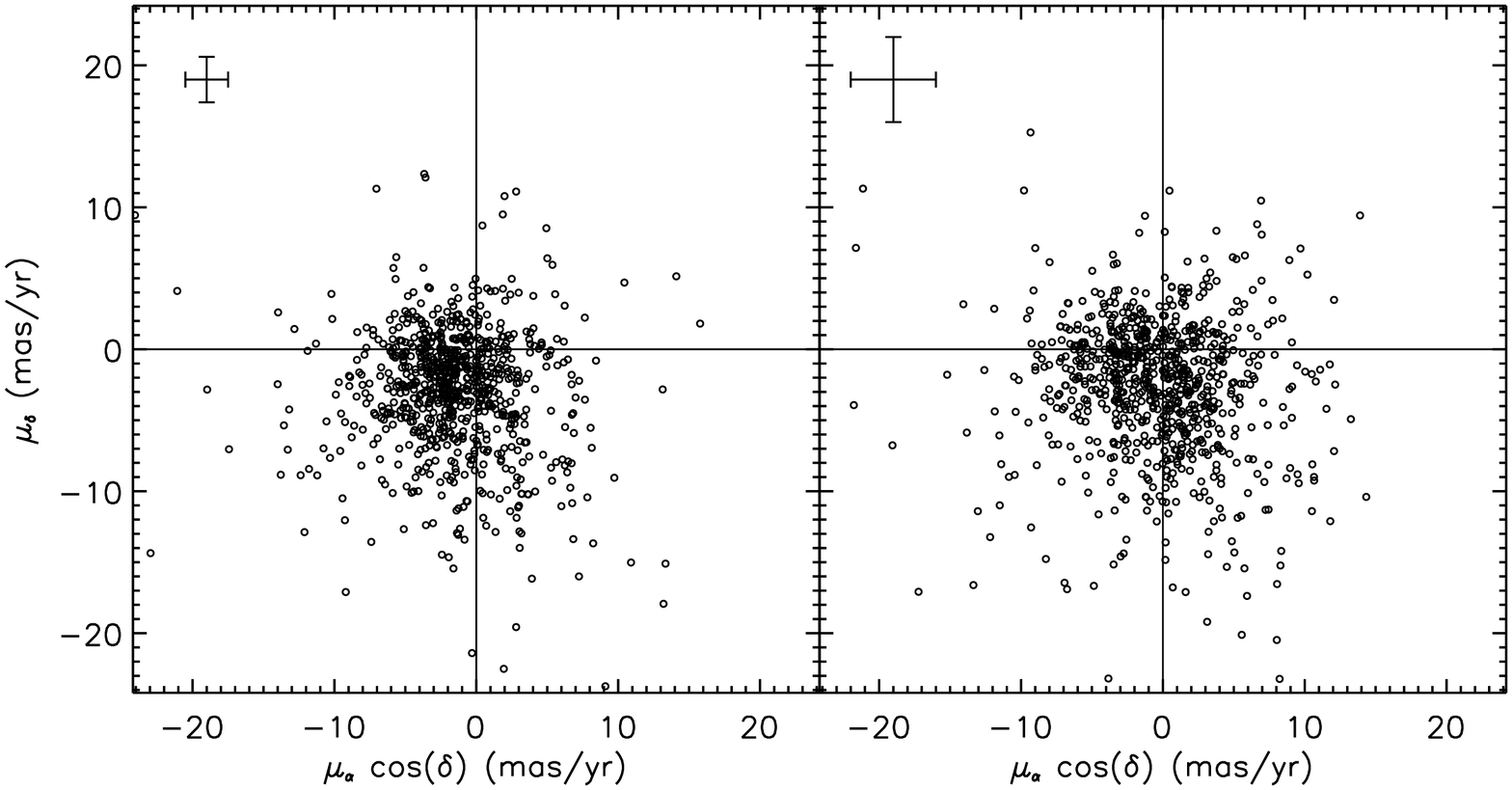}
\caption{{\it Upper panels: } Proper motions of all objects identified as galaxies by SDSS that
appear in our SA 76 dataset and have SDSS/USNO-B proper motions in the
catalog derived by \citet{mml+08}. The left panel shows our
measurements, and the right panel the SDSS PMs; in each plot, a
representative error bar depicts the mean uncertainty for the PM
measurements. The SDSS/USNO-B proper motions are given to integer
precision in the catalog of \citet{mml+08}; for plotting purposes
only, we have applied random shifts of between $\pm$0.5 mas yr$^{-1}$
in each dimension before plotting the points, in order to
differentiate overlapping points. Sigma-clipped (at 2.5$\sigma$) mean
PMs from these distributions give $<(\mu_{\alpha}$ cos
$\delta,\mu_{\delta})>_{\rm SA76} = (-0.12, -0.30)\pm(0.39,0.33)$ mas
yr$^{-1}$ and $<(\mu_{\alpha}$ cos $\delta,\mu_{\delta})>_{\rm SDSS}
= (0.09, 0.66)\pm(0.55,0.47)$ mas yr$^{-1}$. These determinations are
consistent in $\mu_{\alpha}$ cos $\delta$ within the uncertainties,
suggesting that no global offsets are present in that dimension of the
proper motions. However, there is a $\sim2\sigma$ offset in
$\mu_{\delta}$; because the precision of our measurement is superior
to that of the SDSS determination, we choose to retain our proper
motion zero point rather than offsetting to the SDSS/USNO-B frame.
{\it Lower panels: } As in the upper panels, but showing proper
motions of all faint ($g>17$), blue ($0.0<g-r<0.8$) stars (excluding
galaxies) in common between our SA 76 catalog and SDSS.
The distributions appear similar, with less scatter in our more
precise measurements compared to the SDSS
data. \label{fig:sa76_sdss_pms}}
\end{figure}

\begin{figure}%[!htp]
%\epsscale{0.85}
%\plotone{acsmembers_sa76_vs_sdss_vpd_errorbars.v2.eps}
\plotone{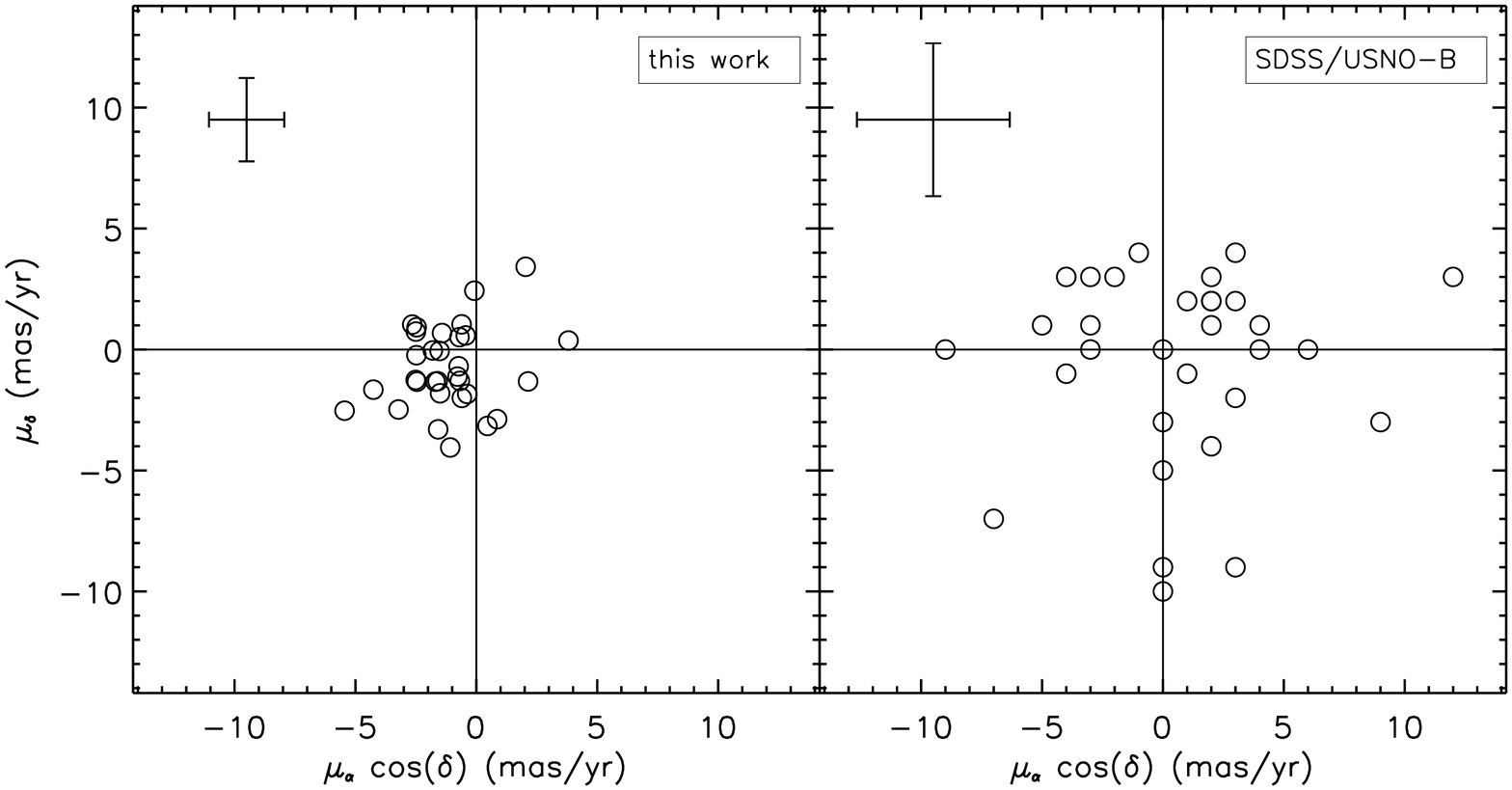}

\caption{Proper motions of the final sample of 31 ACS candidates. The left panel shows our measurements; the right panel those from SDSS/USNO-B. Error bars in each panel represent the typical PM uncertainty. The SDSS proper motions are clearly inconsistent with our measurement for ACS debris: $<(\mu_{\alpha}$ cos $\delta,\mu_{\delta})>_{\rm SA76} = (-1.20,-0.78)\pm(0.34,0.36)$ mas yr$^{-1}$; in fact, from the SDSS PMs, it does not appear that this set of faint stars share a common motion.  \label{fig:acs_sa76_sdss_pms}}

\end{figure}

\subsubsection{Direct Comparison to SDSS/USNO-B Proper Motions}

Another means of determining whether systematic offsets are present in
our data is to compare individual proper motion measurements between
our catalog and another from the literature. The SDSS/USNO-B proper
motions \citep{mml+08} were chosen for this comparison because (a) SA
76 is within the SDSS footprint, and should thus have SDSS/USNO-B
proper motions available for most objects in the field, and (b) the
SDSS/USNO-B proper motions were used by \citetalias{gcm08} to
constrain the orbit of SA 76 using their smaller sample of radial
velocity members. To look for systematics in the zero point used to
fix the absolute reference frame, individual absolute proper motions
were compared for all objects in common between the catalogs that were
identified as galaxies in SDSS (upper panels of
Figure~\ref{fig:sa76_sdss_pms}). The two upper panels show our
measurements on the left and the SDSS/USNO-B PMs on the right, with
the size of a typical uncertainty given by the error bar in the upper
left corner of each plot. Because the SDSS/USNO-B proper motions are
given by \citet{mml+08} to integer precision, many of the points will
overlap each other in such a plot. To make it easier to differentiate
points in the figure, we have applied random shifts of $|\Delta\mu|
\leq 0.5$ mas yr$^{-1}$ in each dimension to the SDSS PMs. Our
measurements are more tightly clumped than the SDSS data, suggesting
that our results are more reliable for these distant,
zero-proper-motion galaxies.  The mean proper motions from these
distributions (clipped at 2.5$\sigma$) are $<(\mu_{\alpha}$ cos
$\delta, \mu_{\delta})>_{\rm SA76} = (-0.12, -0.30)\pm(0.39,0.33)$ mas
yr$^{-1}$ and $<(\mu_{\alpha}$ cos $\delta, \mu_{\delta})>_{\rm SDSS}
= (0.09,0.66)\pm(0.55,0.47)$ mas yr$^{-1}$. The mean SDSS/USNO-B
proper motion of galaxies is inconsistent with zero PM by 1.4$\sigma$
in $\mu_{\delta}$, while our measurements are consistent with zero in
both dimensions. The galaxies' mean $\mu_{\alpha}$ cos $\delta$ from
the two surveys agree within the uncertainties, suggesting that the
difference between our measured $\mu_{\alpha}$ cos $\delta$ for ACS
debris and the expected motion is not due to a zero-point
uncertainty. Note, however, that the SDSS value of $\mu_{\delta}$ for
galaxies differs by more than $2\sigma$ from our determination;
because the precision of our measurements (as reflected by the
individual PM uncertainties as well as the errors in the
sigma-clipped means) is much better than that given by SDSS, we choose
to retain our zero point rather than shift to the SDSS frame.

The lower panels of Figure~\ref{fig:sa76_sdss_pms} are similar to the
upper panels, but compare proper motions for faint ($g>17$), blue
($0.0<g-r<0.8$) stars (excluding galaxies) in common between our SA 76
catalog and SDSS. The distributions are similar, but our proper
motions have roughly twice the precision of the SDSS/USNO-B
determinations. This becomes even clearer when comparing only the
final sample of 31 stars we have selected as ACS candidates
(Figure~\ref{fig:acs_sa76_sdss_pms}). From these 31 stars, our data
yield a proper motion with $\sim0.35$ mas yr$^{-1}$ precision in each
dimension; the SDSS mean PMs are constrained to $\sim0.75$ mas
yr$^{-1}$ in each direction, and produce a mean $\mu_{\alpha}$ cos
$\delta$ = 0.91$\pm$0.77 mas yr$^{-1}$ that differs by $\sim6\sigma$
from our determination. Clearly the SDSS/USNO-B proper motions do not
have the requisite precision at faint magnitudes to constrain distant
tidal stream kinematics.

\subsection{Is Sagittarius Tidal Debris Present Among the ACS Candidates?}

Another stellar population that may be present along the SA 76 line of
sight is leading tidal debris from the Sagittarius (Sgr) dSph.  The
Sgr stream is prominently visible $\sim$5$\arcdeg$ from the position
of SA 76 in Fig. 2 of \citetalias{gcm08} and our
Figure~\ref{fig:starcounts_orbit}.  To determine the expected
properties of Sgr debris in this region, we turn to the recent
comprehensive model of the Sgr tidal debris system from \citet{lm10}.
Figure 9 of this work shows the expected positions, radial velocities,
and distances for the Sgr stream in the SDSS footprint.  SA 76, at
$(\alpha,\delta)_{2000} = (125.3\arcdeg, 14.7\arcdeg)$, is on the
periphery of the Sgr leading debris tail.  All stars within
$\pm$5$\arcdeg$ (in both $\alpha$ and $\delta$) of the position of SA
76 were selected from the best-fit model of \citet{lm10}, and their
kinematical properties compared to stars in SA 76.  The median
expected radial velocity for Sgr leading arm debris thus selected is
$V_{\rm helio,Sgr}$ = -47 km~s$^{-1}$ ($V_{\rm GSR,Sgr}$ = -143
km~s$^{-1}$), with a dispersion of $\sim$15 km~s$^{-1}$.  This is well
outside the ACS candidate velocity selection we have used, so we
expect no contamination from Sagittarius debris in our ACS sample
(note also that there is no clear excess at the Sgr debris velocity in
Figure~\ref{fig:vhel_hist}, so we don't seem to be sampling much, if
any, Sgr leading debris in SA 76).  In addition, \citealt{lm10}
predicts Sagittarius leading arm debris to be at a distance of
$\sim$15-20 kpc in this region, which would place the main sequence
turnoff of Sgr debris $\sim$0.5-1.0 magnitudes fainter than the MSTO
seen in SA 76.

\subsection{Is the EBS Associated With the Anticenter Stream?}

The ACS orbit fit by \citetalias{gcm08} passes through the overdensity
at $(\alpha, \delta)_{2000} \sim (134\arcdeg, 3\arcdeg)$, dubbed the
Eastern Banded Structure (or EBS) by \citetalias{g06a}, on a
subsequent orbital wrap, leading \citetalias{gcm08} to suggest that
the ACS and EBS may be associated.  In \citetalias{gcm08} and
\citetalias{g06a}, a large, $\sim$5-degree gap in the coverage of SDSS
DR5 passed between the EBS and the Anticenter Stream.  Since the
publication of \citetalias{gcm08}, SDSS Data Release 7 (DR7) has been
made public.  We have used this more complete photometric catalog to
generate new filtered star-count maps of the Sloan footprint
(Figure~\ref{fig:starcounts_orbit}).  With the gap now filled in DR7,
the EBS overdensity does not appear to continue west toward the main
ACS.  This may be because the stream associated with the EBS is
curving or inclined away from us, and there are simply fewer stars
contributing to the starcounts at fainter magnitudes.  It could also
be that the portion of the stream to the west of the EBS is "clumpy",
and there is little debris between SA 76 and the main EBS feature.  Of
course, the overdensity at $\sim (134\arcdeg, 3\arcdeg)$ may simply be
where the EBS stops.  The orbit derived from the kinematics we
measured in SA 76, if traced backward from SA 76, would pass through
(or near) the EBS, but the discontinuity between SA 76 and the EBS
``blob'' makes it seem unlikely that the EBS is debris at small
angular separation along the same stream as that sampled by our
candidates in SA 76.  However, this does not rule out an association
between the EBS and the whole of the ACS -- a larger-scale, deep
kinematical study would be necessary to assess their possible physical
association.

\begin{figure}%[!htp]
\epsscale{0.65}
\plotone{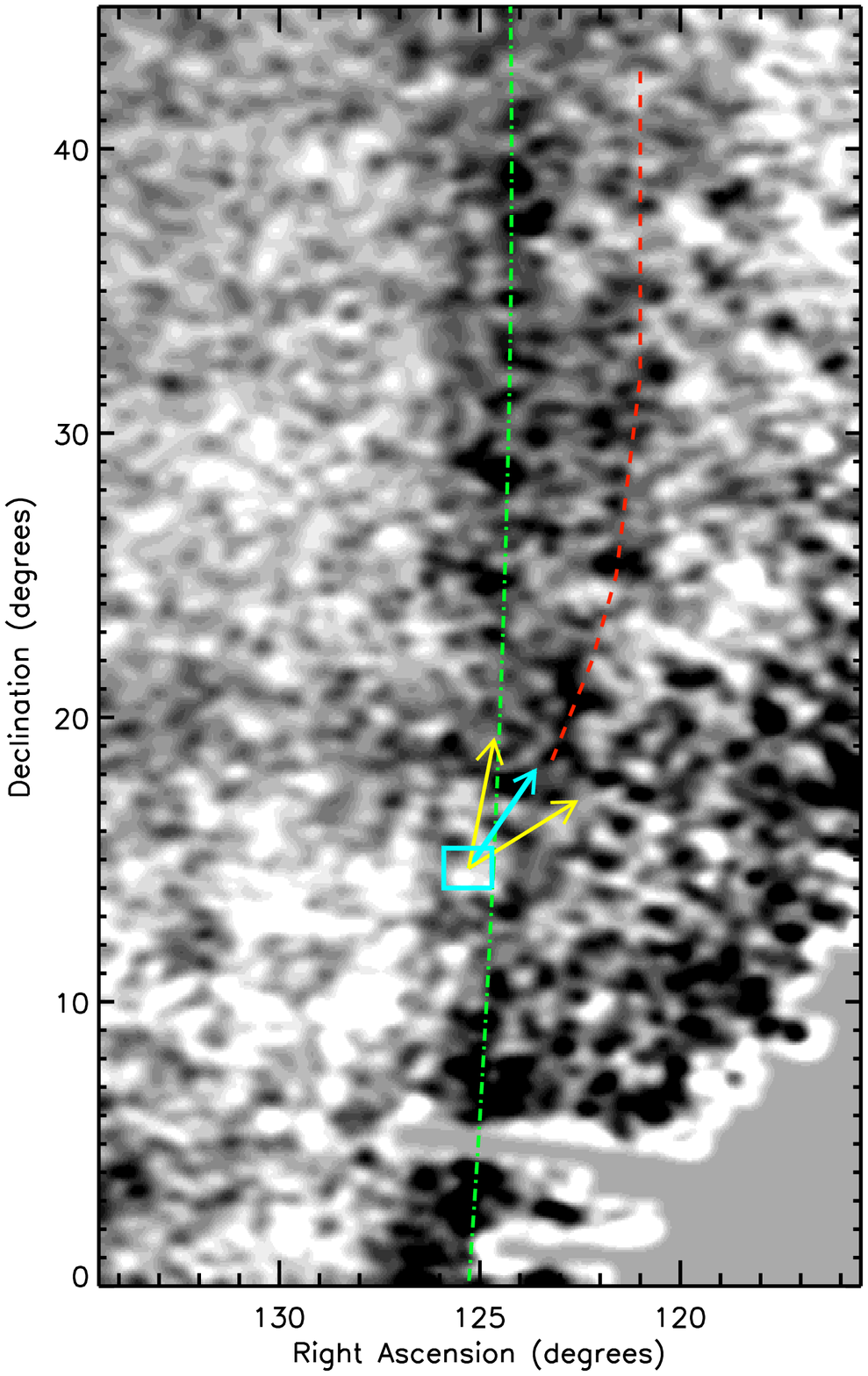}
\caption{Starcount map (as in Figure~\ref{fig:starcounts_orbit}) with
the orbit of \citetalias{gcm08} shown by the (green) dot-dashed line,
and SA 76 denoted by the open square.  A path of the Western substream
possibly traced by ACS debris with kinematics as measured in SA 76 is
highlighted as a (red) dashed curve.  This curve is loosely based on
the peaks identified as the ``West'' substream by \citetalias{g06a},
as well as visual examination of the starcount map and extrapolation
of the proper motion vector. \label{fig:starcounts_weststream}}
\end{figure}

\subsection{Are We Measuring the Motion of a Tributary or Substream of the Main Anticenter Stream?}

\citet{g06a} found that the overdensities of starcounts in the ACS
separate into three distinct components: a broad $\sim2\arcdeg$ - wide
stream running down the center of the system, with two (or more)
narrower streams on the east and west side of the broad component.
Their Figure 2 shows total starcounts as a function of right ascension
in slices at constant declination across the stream.  It is clear from
that figure that there are at least 3 main components, and probably a
number of other sub-streams making up the Anticenter Stream complex.
Thus, in order to measure the bulk motion and derive an orbit for the
stream progenitor, it may be necessary to disentangle these
tributaries and assess the contributions of residual motion about a
common center of mass (i.e. the progenitor dwarf galaxy) that the
separate components may have shared.

The measured motion of ACS candidates in SA 76 (represented as a
vector by the cyan arrow in Figure~\ref{fig:starcounts_orbit}) appears
to follow the western substream studied by \citetalias{g06a}, which
loops outward to the west and separates from the main (or ``central'')
stream component to the north of SA 76 in
Figure~\ref{fig:starcounts_orbit}.  The orientation of our derived ACS
motion from SA 76 stars may represent the peculiar motion of this
substream about the orbital center of the stream progenitor.
Figure~\ref{fig:starcounts_weststream} depicts the same starcount map,
but zoomed to the region near SA 76.  On this figure, we have traced a
possible course of the western substream, which is guided in part by
the peaks identified by \citetalias{g06a} in profiles of starcounts
taken across the stream at positions of constant declination.  This is
simply shown as a guide to the reader to illustrate our suggestion
that the SA 76 kinematics may reflect the motion of the western
tributary.  Note also that the open box denoting the position of SA 76
in Figure~\ref{fig:starcounts_weststream} is in a region of lower
stream density (white in the figure) on the periphery of the ACS.

As mentioned previously, the total 3-D space velocity we measured
($V_{\rm tot} \approx 181$ km s$^{-1}$) differs by only 47 km s$^{-1}$
from the best-fit ACS model motion of \citetalias{gcm08} at the
position of SA 76 ($V_{\rm tot} \approx 225$ km s$^{-1}$).  In testing
the possibility that Pal 12 was previously associated with the
Sagittarius dSph, \citet{dmg+00} calculated an escape velocity of
$\sim$40 km s$^{-1}$ for a globular cluster bound to a Sgr-like 10$^8$
M$_{\odot}$ satellite.  Thus the relative velocities are consistent
with components such as globular clusters or other distinct
kinematical overdensities within the progenitor.  Furthermore, the
angle between these 3-D velocity vectors is only $\sim 27\arcdeg$.
This fact taken together with the similarity of the space velocities
suggests that the orbits, though oriented in slightly different
directions, may be associated, with their slight misalignment a
remnant of peculiar motions about the center of mass of the progenitor
system.  Moreover, the integrals of motion for both orbits are
similar: the total angular momenta (per unit mass) for the two orbits
agree within $\sim$ 20-25\%, and their orbital energies differ by
$\sim$10-15\%.  This is suggests of similar orbits for the main ACS
stream and the motion we have measured in SA 76, in spite of the
misalignment with the visible stream.  It may be that the ACS
represents the remnant of a progenitor satellite that originally had
its own satellites orbiting within its gravitational potential,
similar to the globular clusters associated with the Sgr dSph
(e.g. \citealt{da95,dmg+00,bif+03}) or the Fornax dSph (see, e.g.,
\citealt{fmm+00}).  It is also possible that this richly substructured
stream complex is the result of an infalling group of associated
satellites that fell into the Milky Way together, as in, e.g.,
\citet{lh08} or \citet{dl09}.

The low-inclination, nearly-circular orbit derived for the ACS
suggests that one may expect debris from multiple wraps of the
best-fit orbit (or two separate sub-systems) to be present in SA 76,
perhaps with distinct radial velocity signatures.  Debris from the
subsequent wrap of the orbit in \citetalias{gcm08} passes through the
ACS locus at a distance of $\sim$15 kpc (though at a declination
10-20$\arcdeg$ north of SA 76).  We will consider this as a reasonable
distance for debris from a second orbital wrap.  A difference of 3-5
kpc in distance corresponds to an offset of $\Delta g \sim$ 0.6-1.0
magnitudes in the CMD.  It is unlikely that these two overlapping
stellar populations can be separated (while also separating stream
members from Galactic contamination) without accurate RVs from
medium-to-high resolution spectra for {\it all} stars at the faint end
of our survey.  There is the hint of a concentration of stars blueward
of the obvious ACS main sequence, but below the ACS turnoff (at $g
\sim 21, g-r \sim 0.3$) in the CMD of Figure~\ref{fig:full_cmd} (left
panel), suggestive of a second MSTO.  However, this occurs near the
magnitude limit of the proper motion survey, and we are unable to
explore the possibility of a second population with the data in hand.
We do note that the apparent scatter in proper motions for bright ($g
< $18.5) stars (candidate ACS red giants) discussed in Section 3 could
arise from the superposition of multiple RGBs.  If this is the case,
it would also explain the lack of a narrowly defined RGB among
brighter stars with ACS-like RVs, suggesting that the large PM scatter
in RGB-like stars is a reflection of true differences in PMs of
different populations.  The presence of multiple wraps of stellar
debris in SA 76 could also explain the large metallicity spread seen
in Figure~\ref{fig:feh_hist}.  Many dSphs are known to have radial
gradients in [Fe/H], and, as shown by \citet{cmc+07} for the
Sagittarius tidal tails, the preferential stripping of outer
populations from dSphs gives rise to metallicity gradients along the
debris streams.  If the ACS arises from such a scenario, the
superposition of two populations from multiple orbital wraps (thus
having different mean metallicities) in SA 76 could then produce a
broader metallicity distribution than expected from a single
population.

\subsection{Conclusion: the Origin of the Unexpected Kinematics in SA 76}

Though the reason our measured kinematics in SA 76 are not oriented
along the visible Anticenter Stream remains unclear, we have ruled out
systematic proper motion errors and Sagittarius tidal debris as the
origin of the misalignment.  The space velocity we have measured for
ACS candidates in SA 76 is similar to that of the \citetalias{gcm08}
orbital fit, is directed along a visible substream, and has motion
relative to the main stream similar to what is seen for globular
clusters that orbit other Local Group dwarf galaxies (either intact or
disrupted).  Based on these properties, we argue that our measurements
reflect the motion of a tributary of the main ACS system.  This
substream may be the remnant of either a globular cluster or companion
dwarf galaxy that was bound to the larger system that was the ACS
progenitor.

\section{Are the ACS and Monoceros Related?}

There have been numerous studies of stellar systems in the anticenter
and outer Galactic disk region all purporting to be Monoceros (also
known as "GASS") debris
\citep{nyr+02,cmr+03,iil+03,yng+03,wba+05,vz06,cll+07}.  In this
section, we attempt to sort out the many detections and determine
(based on our measured kinematics) whether the distinct ACS feature is
related to Mon.  The many purported Mon detections within the western
SDSS footprint are highlighted as various colored symbols in the
starcount map of Figure~\ref{fig:starcounts_orbit}.  From this figure,
it is clear that Mon, as reported in previous work, would be a large
feature spanning much of the low-latitude sky near the Galactic
anticenter.  The ACS, on the other hand, was eventually revealed to be
a narrow, well-defined stellar stream (with even narrower substreams)
by \citetalias{g06a}.  A few of the claimed Mon detections appear to
be spatially coincident with the ACS (e.g. the points from
\citealt{wba+05}, one of the \citealt{nyr+02} fields, the
\citealt{iil+03} point, and a handful of the M giants from
\citealt{cmr+03}), while the majority of the remaining Mon candidates
are at lower latitudes than the ACS.  It is possible that the two
systems are associated, but if they are distinct systems, some of the
stars and overdensities previously associated with Mon may actually be
members of the Anticenter Stream.  Accurate kinematical and stellar
abundance study of large numbers of stars in this region (preferably
covering a contiguous area) may make it possible to determine whether
these are associated features of the same disruption event, a spatial
coincidence of unrelated systems, or confusion of not well
discriminated early studies of structures near the Galactic antienter.
In this section, we compare the properties of Monoceros and the ACS to
explore their possible association.

\subsection{Metallicities}

The abundance spread for claimed Monoceros detections is not fully
understood, with a variety of conflicting results making this feature
difficult to understand.  One of the earliest studies of Monoceros
\citep{yng+03} estimated [Fe/H] = -1.6 $\pm$ 0.3 from SDSS spectra of
MSTO F stars, while \citet{cmr+03} found a much higher [Fe/H] = -0.4
$\pm$ 0.3 for 2MASS-selected M giants.  From these two results alone,
it is obvious that if the various Mon overdensities derive from a
single structure, then there must have been a significant metallicity
spread within the Mon progenitor.  \citet{isj+08} derive [Fe/H] =
-0.95 for Monoceros based on photometric metallicites of thousands of
F- and G-type dwarfs near the Galactic anticenter from the SDSS
database, and also show that this metallicity is quite distinct from
the MW halo and disk in this direction (see their Figure 18).  Our
result for the ACS is generally rather similar to these results for
Monoceros, in that we find a predominantly metal-poor population, but
with broad scatter.  Additional work on A/F-type stars by
\citet{wba+05} found $<$[Fe/H]$>$ = -1.37, with scatter of $\sim$0.5
dex -- both their mean metallicity and the spread of measurements are
comparable to what we have found for the ACS from stars of similar
spectral type.  The stars in this \citet{wba+05} study are located in
two SDSS plates that are on or near the obvious ACS feature in
Figure~\ref{fig:starcounts_orbit} (where \citealt{wba+05} fields are
shown as open red circles).  The population identified as Mon debris
in the southernmost of these fields (Plates 1149/1154, at ($\alpha,
\delta) \sim (125.1, 2.7)$) had measured radial velocity of 88 km
s$^{-1}$, which is consistent with the RV for the ACS orbit derived by
\citetalias{gcm08} at that position.  Thus, though \citet{wba+05} have
claimed these to be Monoceros stars, it is likely that their study is
sampling ACS debris from the same populations we are studying in SA
76.

Finally, we note that \citet{cmc+10} show that many M giants (selected
from \citealt{cmr+03}) in their high-resolution spectroscopic study
near the anticenter have [Fe/H]$\sim$ -0.9 (as expected from
\citealt{isj+08}), with a tail extending to higher metallicities
(confirming the results of \citealt{cmr+03}).  This work shows that
these Mon M giants have $\alpha$ and s-process elemental abundance
trends similar to those of M giants in the Sagittarius tidal tails, as
well as other dSphs.  While this suggests a dwarf galaxy origin for
the Monoceros stars, it cannot definitively rule out that Mon is the
result of a ``puffing up'' of the disk due to an encounter, since the
outer disk itself may be formed hierarchically from merger events.
The similarity of the rather narrow Anticenter Stream (which is likely
the result of a dSph disruption) to the much more broadly distributed
Monoceros feature could be because the ACS is the coherent remnant of
a larger disruption event that produced the entire Mon "ring".
Alternatively, it may simply be that spatial coincidence has led to
the conflation of two distinct structures, and that confusion leads to
ACS members "contaminating" determinations of Mon properties.

\begin{figure}%[!htp]
\epsscale{0.65}
\plotone{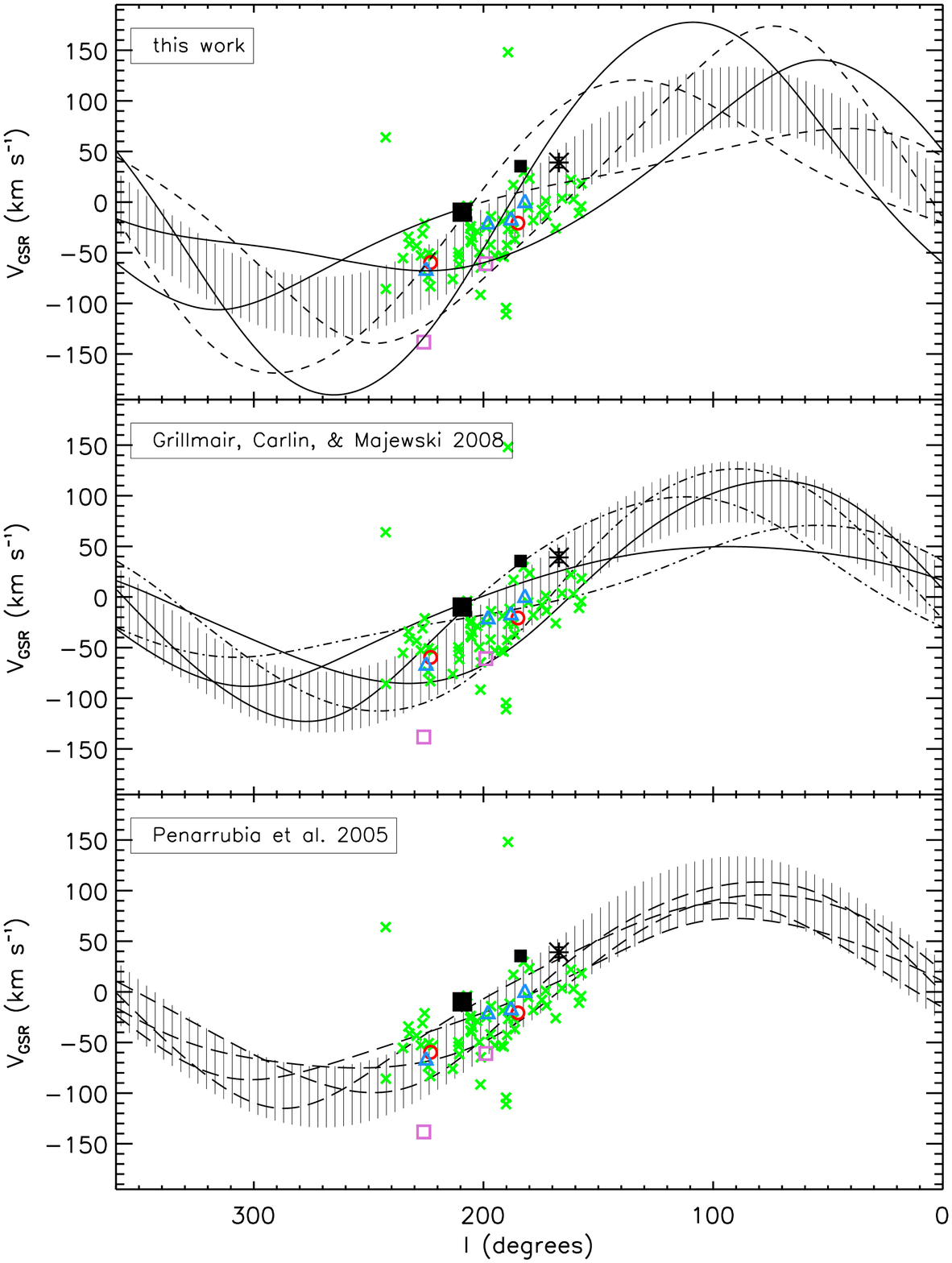}
\caption{Galactocentric radial velocities as a function of Galactic
longitude for our derived orbit (top panel; solid line: orbit
integrated backward from SA 76 for $\sim$1 Gyr, dashed line: forward
integration), the \citet{gcm08} orbit (middle panel; solid line:
backward integration, dot-dash: forward), and \citet{pmr+05} orbit
(bottom; long-dash).  In each panel, a hatched region centered on the
circular velocity curve is shown, as derived by \citet{cmr+03} to fit
their velocities of candidate Monoceros M giants: $V_{circ}$ = 220 km
s$^{-1}$ at Galactocentric distance of $R_{GC}$ = 18 kpc, with
velocity dispersion of 30 km s$^{-1}$.  Also shown in each panel are
Monoceros detections from the literature (symbols and colors as in
Figure~\ref{fig:starcounts_orbit}).  The large asterisk shows the mean
radial velocity for purported Monoceros debris in SA 71, a field in
which \citet{ccg+08} also measured 3-D kinematics.  The error bar for
this point is $\pm$5 km s$^{-1}$, which is smaller than the size of
the plotted point.  A large filled square shows our measured velocity
for ACS debris at the position of SA 76 and the smaller filled square
represents the velocity of ACS debris in ACS-B
(\citetalias{gcm08}). \label{fig:lv_compare}}
\end{figure}

\begin{figure}%[!htp]
\epsscale{0.7}
\plotone{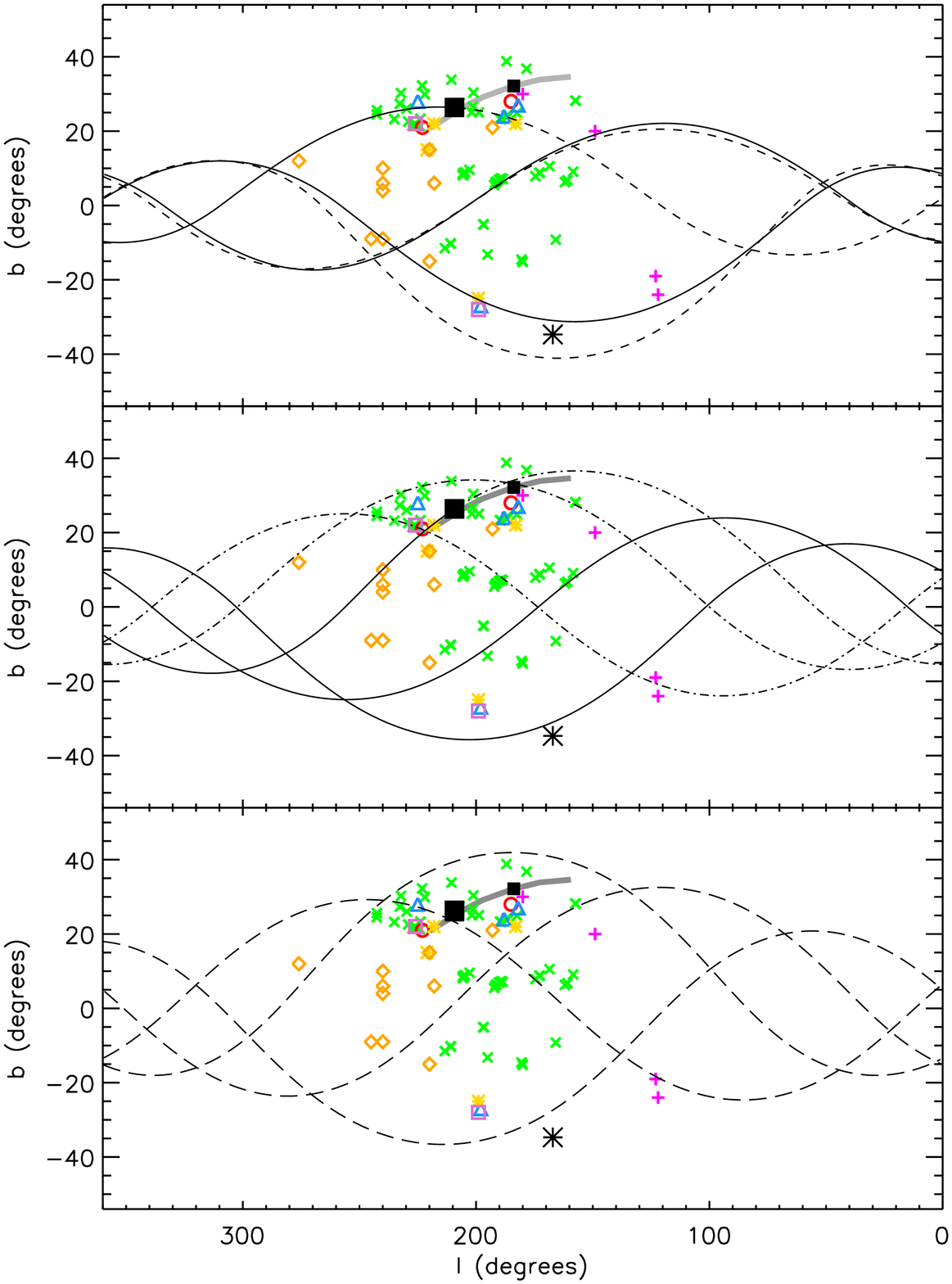}
\caption{Orbital paths in Galactic coordinates, with panels and
symbols as in Figure~\ref{fig:lv_compare}. The grey filled ``bars'' in
each panel represent the area covered by ACS debris in the western
portion of the SDSS footprint. \label{fig:lb_compare}}
\end{figure}

\subsection{Kinematics}

Predicted Galactocentric radial velocities as a function of Galactic
longitude are shown in Figure~\ref{fig:lv_compare} for our derived
orbit of SA 76 ACS debris (top panel), with the best-fit orbits from
\citetalias{gcm08} (ACS) and \citet[Mon]{pmr+05} in the lower two
panels.  A similar plot in Figure~\ref{fig:lb_compare} shows the
spatial distribution of the ACS and Monoceros orbits in Galactic
coordinates.  Comparison data from the literature are shown in both
figures using the same colors and symbols as in
Figure~\ref{fig:starcounts_orbit} for claimed Monoceros detections
near our SA 76 field.  The large filled square representing our
measured velocity for ACS debris at the position of SA 76 is slightly
outside the hatched region in Figure~\ref{fig:lv_compare}, which
represents an object in a circular orbit with velocity of 220 km
s$^{-1}$ at R$_{GC}$ = 18 kpc, with a dispersion of 30 km s$^{-1}$
(similar to \citealt{cmr+03}; note that, assuming a flat MW rotation
curve, this velocity and dispersion are thin disk-like).  Our newly
derived orbit is the only one of the three shown that exhibits
significantly non-circular motion (i.e. extends well outside the
hatched region) in Fig.~\ref{fig:lv_compare}.  The orbit derived by
\citet{pmr+05} for Monoceros predicts a velocity in SA 76 that is
similar to what we have measured, but cannot reproduce the spatial
distribution of the ACS, as noted in \citetalias{gcm08}.  We note,
however, that our orbit derived from kinematics in SA 76 also fails to
reproduce the swath of ACS debris (the grey "bar" in
Figure~\ref{fig:lb_compare}) that spans the entire declination range
of the western SDSS footprint.  Our derived orbit traverses slightly
higher Galactic latitudes in Figure~\ref{fig:lb_compare} than the
other two models, but is qualitatively quite similar to the result of
\citetalias{gcm08}, which suggests that our current study is measuring
distinct ``tributaries'' or successive orbital wraps of the same
stream system modeled in \citetalias{gcm08}.  The \citetalias{gcm08}
orbit also reproduces many of the known detections of Mon debris in
both velocity and position, but further detailed study incorporating
all ACS and Mon data would be necessary to verify an association of
the ACS with the Monoceros ring.

With 3-D kinematics, we can assess the similarity of the orbits
derived by various studies by examining the integrals of motion.  As
discussed in \S6, the mean motion we have measured in SA 76
corresponds to an angular momentum (per unit mass) of $L_Z \sim
-2960\pm330$ kpc km s$^{-1}$, placing the SA 76 ACS debris on a
prograde orbit.  This differs by only $\sim20\%$ from the $L_Z \sim
-3600$ kpc km s$^{-1}$ predicted by the \citetalias{gcm08} ACS orbit
along the SA 76 line of sight.  \citet{ccg+08} measured
three-dimensional stellar kinematics from an overdensity of purported
Monoceros debris in SA 71 and found $L_Z \sim -3140\pm460$ kpc km
s$^{-1}$, which agrees with the ACS results within the uncertainties.
The orbit derived from our kinematics in SA 76 produces a radial
velocity near the \citet{ccg+08} measurement for SA 71 at $l \approx
167\arcdeg$ (Fig.~\ref{fig:lv_compare}), and also passes near the
spatial location of SA 71 (Fig.~\ref{fig:lb_compare}), so it is
possible that the debris in these two fields originated from the same
progenitor.  \citet{isj+08} studied kinematics near the anticenter
using thousands of stars from the SDSS database at Galactocentric
distances $13 < R_{\rm GC} < 16$ kpc, and found an excess of stars
with very little vertical (i.e., along Galactic latitude,
perpendicular to the disk) motion and rotation velocities faster than
the local standard of rest (LSR) rotation by $\sim20-50$ km s$^{-1}$.
At a distance of 15 kpc, this results in $-4000 < L_Z < -3500$ kpc km
s$^{-1}$, which is also similar to the angular momenta from both the
ACS results and the Mon debris in SA 71.  We note, though, that in our
result, the \citetalias{gcm08} orbit, and the SA 71 measurement
\citep{ccg+08}, the rotation component of the motion lags the LSR,
while \citet{isj+08} find Mon to be {\it faster} than the LSR by
$\sim50$ km s$^{-1}$.  Ultimately, while all of these measurements are
finding similar orbital angular momenta, there is still insufficient
discriminating power in the results to assess whether the ACS and Mon
share a common origin.

It is important to remind the reader that the derived orbit we have
shown was integrated in a spherical Milky Way halo.  The best fits for
Monoceros from \citet{pmr+05} require an oblate dark matter halo
(i.e. $q <$ 1, where $q$ is the ratio of minor axis to the major axis
of the potential) to reproduce the observed characteristics of the
detections assumed to be part of the Monoceros system at the time.
The difference in halo potentials used may be the source of at least
some of the discrepancy between their model and our derived orbits in
Figures~\ref{fig:lv_compare} and \ref{fig:lb_compare}.  Constraints on
the halo flattening based on the Sagittarius dSph tidal stream have
variously argued that the halo is prolate ($q >$ 1; \citealt{h04}),
nearly spherical ($q =$ 0; \citealt{ili+01,fbe+06}), or slightly
oblate \citep{jlm05,ljm05,mga+04}.  More recently, \citet{lmj09} and
\citet{lm10} were able to simultaneously reproduce most observed
characteristics of the Sagittarius stream by modeling the halo as
triaxial, thus reconciling the prior seemingly conflicting results,
wherein a prolate halo was required to match radial velocities of
leading debris, while an oblate halo was needed to get the correct
positions along the leading arm.  Further constraints on the MW halo
shape can be derived by modeling the precession of other tidal
streams, and once the ACS and Monoceros systems are well understood,
they might be used to trace the low-latitude dark matter structure of
the Milky Way.  However, for this work we chose simply to use a
spherical halo model to assess the general qualitative structure of
the orbit implied by the 3-D velocity we have measured.  A more
comprehensive study utilizing all extant data for the ACS and Mon
features (which also would require sorting out which data to include
for each system) and varying halo parameters is beyond the scope of
this work.  We simply note here that changing the halo shape (and
other parameters, such as the rotation speed at the solar circle, or
the distance of the Sun from the Galactic center) would alter the
exact characteristics of the orbit shown in
Figures~\ref{fig:orbit_xyz}, \ref{fig:lv_compare}, and
\ref{fig:lb_compare}, but not the general structure of a low-latitude,
ring-like feature.

\section{Summary}

We have found that the 3-D kinematics of SA 76 stars selected to be
Anticenter Stream members produce unexpected results -- the orbit
derived from the measured motions does not follow the obvious stream
(and the orbit fit to it by \citetalias{gcm08}) from the SDSS
database, but rather is inclined to it by $\sim$30$\arcdeg$.  This
makes it difficult to reach conclusions about the global structure of
the ACS based on our measurements -- we are unable to confirm or rule
out an association of the ACS with the many Monoceros stream
detections or to speculate on a possible progenitor for the system.
We do conclude, however, that our findings can best be explained if
the ACS debris in SA 76 is part of one of the apparently kinematically
cold substreams or tributaries found by \citet{g06a} to make up the
larger ACS stream.  In fact, the measured motions of SA 76 ACS debris
and the ACS orbit of \citet{gcm08} suggest that the substreams may be
the remnants of two (or more) satellites that fell into (and were
disrupted by) the Milky Way together.  More detailed photometric and
chemodynamical studies of the ACS are necessary to explore this
possibility.  In addition, the association of the Eastern Banded
Structure with the ACS could possibly be confirmed by studies
extending beyond the $\delta\sim0\arcdeg$ limit of the SDSS footprint.

We gratefully acknowledge support by NSF grant AST-0807945 and
NASA/JPL contract 1228235.  D.I.C. acknowledges the support of NSF
grant AST-0406884.

{\it Facilities:} \facility{WIYN (Hydra)}, \facility{Sloan}

Funding for the SDSS and SDSS-II has been provided by the Alfred
P. Sloan Foundation, the Participating Institutions, the National
Science Foundation, the U.S. Department of Energy, the National
Aeronautics and Space Administration, the Japanese Monbukagakusho, the
Max Planck Society, and the Higher Education Funding Council for
England. The SDSS Web Site is http://www.sdss.org/.

The SDSS is managed by the Astrophysical Research Consortium for the
Participating Institutions. The Participating Institutions are the
American Museum of Natural History, Astrophysical Institute Potsdam,
University of Basel, University of Cambridge, Case Western Reserve
University, University of Chicago, Drexel University, Fermilab, the
Institute for Advanced Study, the Japan Participation Group, Johns
Hopkins University, the Joint Institute for Nuclear Astrophysics, the
Kavli Institute for Particle Astrophysics and Cosmology, the Korean
Scientist Group, the Chinese Academy of Sciences (LAMOST), Los Alamos
National Laboratory, the Max-Planck-Institute for Astronomy (MPIA),
the Max-Planck-Institute for Astrophysics (MPA), New Mexico State
University, Ohio State University, University of Pittsburgh,
University of Portsmouth, Princeton University, the United States
Naval Observatory, and the University of Washington.

\bibliographystyle{apj}
%\bibliography{bibdesk_library}

\end{document}